\documentstyle[aaspp4,12pt,tighten,eqsecnum,flushrt,axodraw]{article}

\slugcomment{FERMILAB-Pub-96/023-A}

\begin{document}
\def\la{\mathrel{\mathpalette\fun <}}
\def\ga{\mathrel{\mathpalette\fun >}}
\def\fun#1#2{\lower3.6pt\vbox{\baselineskip0pt\lineskip.9pt
        \ialign{$\mathsurround=0pt#1\hfill##\hfil$\crcr#2\crcr\sim\crcr}}}

\title{Loop Corrections in Non-Linear Cosmological Perturbation Theory 
II. Two-point Statistics and Self-Similarity}

\author{ Rom\'an Scoccimarro$^{1,2}$ and Joshua A. Frieman$^2$\altaffilmark{,3}}
\affil{${}^1$Department of Physics and Enrico Fermi Institute, 
University of Chicago, Chicago, IL 60637}
\affil{${}^2$NASA/Fermilab Astrophysics Center, Fermi National Accelerator 
Laboratory, P.O.Box 500, Batavia, IL 60510}
\altaffiltext{3}{Also Department of Astronomy 
and Astrophysics, University of Chicago, Chicago, IL 60637}
\date{\today}

\begin{abstract} 
We calculate the lowest-order non-linear 
contributions to the power spectrum, 
two-point correlation function, and smoothed variance of the density field, 
for Gaussian initial conditions and scale-free initial power spectra, 
$P(k) \sim k^n$. These results extend and in some cases correct previous 
work in the literature on cosmological perturbation theory.  
Comparing with the scaling behavior 
observed in N-body simulations, we find that the validity of non-linear 
perturbation theory depends strongly on the spectral index $n$. 
For $n<-1$, we find excellent agreement over scales where 
the variance $\sigma^2(R) \la 10$; however,  
for $n \geq -1$, perturbation theory predicts deviations from 
self-similar scaling (which increase with $n$)  
not seen in numerical simulations. This anomalous scaling suggests that 
the principal assumption underlying cosmological perturbation theory, 
that large-scale fields can be described perturbatively 
even when fluctuations are highly 
non-linear on small scales, breaks down beyond leading order 
for spectral indices $n \geq -1$. For $n < -1$, the power spectrum,  
variance, and correlation function in the  scaling regime 
can be calculated using dimensional regularization.

\end{abstract}


\section{Introduction} 
\label{sec:intro}

Several independent arguments suggest that the growth of 
cosmological density perturbations on large scales can be 
described by perturbation theory, even when the density and 
velocity fields are highly non-linear on small scales. For 
example, analytic and numerical work showed that linear 
perturbation theory describes the evolution of the large-scale 
density power spectrum $P(k)$, provided the initial spectrum 
falls off less steeply than $k^4$ for small 
$k$~(\cite{Zel'dovich65,Peebles74,PeGr76,Peebles80}).  In addition, for 
Gaussian initial conditions, 
leading-order non-linear perturbative calculations of higher 
order moments of the density field, e.g., the skewness and 
kurtosis, agree well with 
N-body simulations in the weakly non-linear regime, where 
the variance of the smoothed density field $\sigma^2(R) 
\equiv \langle \delta^2(R) \rangle \la 0.5 -  
1$~(Juszkiewicz, Bouchet \& Colombi 1993,
\cite{Bernardeau94c,GaBa95},
Baugh, Gazta\~naga \& Efstathiou 1995).  
Thus, leading-order perturbation theory has been 
shown to work surprisingly well in the cases where comparisons 
with numerical simulations have been made.

The success of leading-order cosmological perturbation theory 
raises questions: since the perturbation series is most likely 
asymptotic, 
what happens beyond leading order---does the agreement with 
simulations improve or deteriorate? More generally, what sets the limits of 
perturbation theory, beyond which it breaks down? These 
questions have become more urgent since it has been shown that 
leading-order perturbation theory appears to 
provide an adequate description even on scales where {\it 
next-to-leading order} and higher order perturbative 
contributions would be expected to become important.

To address these questions, one must calculate loop 
corrections, {\it i.e.}, corrections {\it beyond} leading order, in 
non-linear cosmological perturbation theory (NLCPT). This is the 
second paper of a series devoted to this topic. In the 
first paper~(\cite{ScFr96}), we applied diagrammatic techniques to 
calculate loop corrections to 1-point cumulants of unsmoothed 
fields, such as the variance and skewness, for Gaussian initial 
conditions.  Here, we calculate the 1-loop (first 
non-linear) corrections to the power spectrum, the 
volume-averaged two-point correlation function, and the variance 
of the smoothed density field for scale-free initial spectra, 
$P(k) \sim k^n$.  While the linear power spectrum for the Universe 
is not scale-free (on both observational and theoretical 
grounds), scale-free spectra are useful approximations over limited 
ranges of wavenumber $k$; they also have the advantage of yielding  
analytic closed form results. 
In a forthcoming paper, we will present 1-loop 
corrections to the bispectrum (the three-point 
cumulant in Fourier space) and the skewness of the smoothed 
density field for scale-free and cold dark matter spectra.

One-loop corrections to the two-point correlation function and 
power spectrum have been previously studied in the literature 
~(\cite{Juszkiewicz81,Vishniac83}, Juszkiewicz, Sonoda \& Barrow 1984,
\cite{Coles90,SuSa91}, Makino, Sasaki \& Suto 1992, \cite{JaBe94,BaEf94}). 
Multi-loop corrections to the power spectrum were considered 
by~\cite{Fry94}, including the full contributions up to 2 loops 
and the most important terms at large $k$ in 3- and 4-loop order.  
Some of our results overlap in particular with the analytic 
results for the 1-loop power spectrum reported 
by Suto \& Sasaki (1991) and  in Makino et al. (1992). 
However, since we found that some of 
their expressions contain errors, we present complete corrected 
expressions for the 1-loop power spectrum.  One-loop corrections 
to the variance were studied numerically for Gaussian smoothing 
by \L okas et al. (1995).  We correct some numerical errors in their 
results, extend them to include top-hat smoothing and to the 
average two-point correlation function, and provide analytic 
derivation of some of the numerical results.  

The limiting 
behavior of the 1-loop corrections on large scales leads us to 
reconsider the issue of self-similarity in perturbation theory
~(\cite{DP77,Peebles80}). It is commonly accepted 
that the evolution of density perturbations from scale-free initial 
conditions in an Einstein-de Sitter universe is statistically 
self-similar. N-body simulations have generally found self-similar 
scaling for $-3 < n < 1$, although the results for $n<-1$ have 
been somewhat ambiguous due to problems of dynamic range in the 
simulations~(\cite{EFAL88,BG91,RG91,G92,CBH95,Jain95,PCOS95}).  
The issue of self-similarity in NLCPT has recently 
been investigated by Jain \& Bertschinger (1995), who present arguments 
that the perturbative evolution is also self-similar.  While we do not 
disagree with their calculations, we start from a different 
premise regarding cutoffs in the initial spectrum, 
which leads us to conclude that loop corrections 
break self-similar scaling if the spectral index $n \geq -1$.  In 
the regime where we do find scaling, $n < -1$, we use dimensional 
regularization to calculate analytically the 
asymptotic behavior of the 1-loop power spectrum, variance, and average 
correlation function; these results agree quite well with the 
`universal scaling' extracted from numerical 
simulations~(\cite{HKLM91,PeDo94}, Jain, Mo \& White 1995). 

The paper is organized as follows.  In Sec.~\ref{sec:dynstat} we 
discuss non-linear perturbation theory and review the 
diagrammatic approach developed in~(\cite{Fry84,GGRW86,ScFr96}) to the 
calculation of statistical quantities.  Section~\ref{sec:results} 
presents the results of 1-loop calculations for the power 
spectrum, the smoothed variance, and average two-point 
correlation function.  Self-similarity in perturbation theory is 
the subject of Sec.~\ref{sec:selfsimpt}.  We discuss the 
conditions under which 1-loop corrections exhibit self-similarity 
and compare our results with the universal scaling 
hypothesis, based on numerical 
simulations.  Section~\ref{conclusions} contains our conclusions.

\section{Dynamics and Statistics} 
\label{sec:dynstat}

\subsection{The Equations of Motion}
\label{sec:eqsmotion}

Assuming the universe is dominated by pressureless dust (e.g., cold dark 
matter), in the single-stream approximation (prior to orbit crossing) one can 
adopt a fluid description of the cosmological N-body problem. In this 
limit, the relevant equations of motion 
correspond to conservation of mass and 
momentum and the Poisson equation~(e.g., \cite{Peebles80,ScFr96}),
\begin{mathletters}
\label{eqsmotionrealspace}

\begin{equation}
	{\partial \delta({\bf x},\tau) \over{\partial \tau}} + \nabla \cdot  \{
	[1+\delta{({\bf x},\tau)}] {\bf v}({\bf x},\tau) \} =  0
	\label{continuity},
\end{equation}
\begin{equation}
     {\partial {\bf v}({\bf x},\tau) \over{\partial \tau}} + {{\cal H}(\tau)}\ 
     {\bf v}({\bf x},\tau) + [{\bf 
	v}({\bf x},\tau) \cdot \nabla] {\bf v}({\bf x},\tau) =  - \nabla 
	\Phi({\bf x},\tau)
	\label{euler},
\end{equation}	
\begin{equation}
	\nabla^2 \Phi({\bf x},\tau)  = {3\over 2} \Omega {\cal H}^2(\tau) 
	\delta{({\bf x},\tau)}
	\label{poisson}
\end{equation}
 \end{mathletters}	 

\noindent Here the density contrast $\delta{({\bf x},\tau)} \equiv
\rho{({\bf x},\tau)}/ \bar \rho  - 1$, with ${\bar \rho}(\tau)$ 
the mean density 
of matter, $ {\bf v} \equiv d{\bf x}/d\tau $ represents the velocity field 
fluctuations about the Hubble flow,  $ {\cal H}\equiv {d\ln a /{d\tau}}$
is the conformal expansion rate,  $a(\tau)$ is the cosmic 
scale factor, ${\bf x}$ denote comoving coordinates, $\tau=\int dt/a$ is 
the conformal time,  
$\Phi$ is the gravitational potential due to the 
density fluctuations, and the density parameter $\Omega = {\bar \rho}/
\rho_c = 8\pi G {\bar \rho} a^2/3{\cal H}^2$. 
Note that we have implicitly assumed the Newtonian 
approximation to General Relativity, valid on scales less than the Hubble
 length $a {\cal H}^{-1}$. We take the velocity field to be 
irrotational, so it can be completely described by its divergence 
$ \theta \equiv \nabla \cdot {\bf v }$. 
Equations~(\ref{eqsmotionrealspace}) hold in an arbitrary 
homogeneous and isotropic background Universe which evolves according to the
Friedmann equations; henceforth, for simplicity we assume an Einstein-de Sitter 
background, $\Omega =1$ (with vanishing 
cosmological constant), for which  $a \propto \tau^2$ and $3 \Omega 
{\cal H}^2 /2 = 6/\tau^2$.

We take the divergence of Equation~(\ref{euler}) and Fourier transform the 
resulting equations of motion according to the convention

\begin{equation}
	\tilde{A}({\bf k},\tau)=\int {d^3x \over{(2 \pi)^3}} \exp (-i {\bf k }
	\cdot {\bf x})\  A({\bf x},\tau)
	\label{fourier},
\end{equation}

\noindent for the Fourier transform of any field $A({\bf x},\tau)$, 
where, here and throughout, ${\bf k}$ is a comoving wavenumber. This yields
	
\begin{mathletters}
\label{eqsfourier}
\begin{equation}
	{\partial \tilde{\delta}({\bf k},\tau) \over{\partial \tau}} + 
	\tilde{\theta}({\bf k},\tau) = - \int d^3k_1 \int d^3k_2 \delta_D({\bf k} 
	- {\bf k}_1 - {\bf k}_2) \alpha({\bf k}, {\bf k}_1) \tilde{\theta}({\bf k}_1,\tau) 
	\tilde{\delta}({\bf k}_2,\tau)
	\label{ddtdelta},
\end{equation}
\begin{eqnarray}
	{\partial \tilde{\theta}({\bf k},\tau) \over{\partial \tau}} &+& 
	{\cal H}(\tau)\ \tilde{\theta}({\bf k},\tau) + {3\over 2}  {\cal 
	H}^2(\tau) \tilde{\delta}({\bf k},\tau) = \nonumber \\
      &-& \int d^3k_1 \int d^3k_2 \delta_D({\bf k} 
	- {\bf k}_1 - {\bf k}_2) \beta({\bf k}, {\bf k}_1, {\bf k}_2) 
	\tilde{\theta}({\bf k}_1,\tau) \tilde{\theta}({\bf k}_2,\tau)
	\label{ddttheta},
\end{eqnarray}
\end{mathletters}	 

\noindent ($\delta_D$ denotes the  three-dimensional Dirac delta distribution),  
where the functions 

\begin{equation}
	\alpha({\bf k}, {\bf k}_1) \equiv {{\bf k} \cdot {\bf 
k}_1 \over{ k_1^2}}, \ \ \ \ \ \beta({\bf k}, {\bf 
k}_1, {\bf k}_2) \equiv {k^2 ({\bf k}_1 \cdot {\bf k}_2 )\over{2 k_1^2 
k_2^2}}
	\label{albe}
\end{equation}

\noindent encode the non-linearity of the evolution (mode coupling) and come from the 
non-linear terms in the continuity equation~(\ref{continuity}) and the Euler 
equation~(\ref{euler}) respectively.

\subsection{Perturbation Theory Solutions}
\label{sec:recrel}

Equations~(\ref{eqsfourier}) are very difficult to solve in general, being coupled integro-differential equations with no small 
parameter. The perturbative approach to the problem is to consider 
perturbations about the linear solution, effectively treating the 
variance of the linear fluctuations as a small parameter.
In this case, Eqs.~(\ref{eqsfourier})  can be formally  solved 
via a perturbative expansion, 

\begin{equation}
	\tilde{\delta}({\bf k},\tau) = \sum_{n=1}^{\infty} a^n(\tau) 
	\delta_n({\bf k}),\ \ \ \ \ \tilde{\theta}({\bf k},\tau) = 
	{\cal H}(\tau) \sum_{n=1}^{\infty} a^n(\tau) \theta_n({\bf k})
	\label{ptansatz},
\end{equation}

\noindent where only the fastest growing mode is taken into account. 
At small $a$,  
the series are dominated by their first terms, and since  $\theta_1({\bf k}) = 
-\delta_1({\bf k}) $ from the continuity equation, $\delta_1({\bf k})$ 
completely characterizes the linear fluctuations. The equations of 
motion~(\ref{eqsfourier}) determine $\delta_n({\bf k}) $ and 
$\theta_n({\bf k})$  in terms of the linear fluctuations,

\begin{mathletters}
\label{solu}
\begin{equation}
	\delta_n({\bf k}) = \int d^3q_1 \ldots \int d^3q_n \delta_D({\bf k} - 
	{\bf q}_1 - \ldots - {\bf q}_n) F_n^{(s)}({\bf q}_1,  \ldots  ,{\bf q}_n) 
	\delta_1({\bf q}_1) \ldots \delta_1({\bf q}_n)
	\label{ec:deltan},
\end{equation}
\begin{equation}
		\theta_n({\bf k}) = - \int d^3q_1 \ldots \int d^3q_n \delta_D({\bf k} - 
	{\bf q}_1 - \ldots - {\bf q}_n) G_n^{(s)}({\bf q}_1,  \ldots  ,{\bf q}_n) 
	\delta_1({\bf q}_1) \ldots \delta_1({\bf q}_n)
	\label{ec:thetan},
\end{equation}
\end{mathletters}


\noindent where $F_n^{(s)}$ and $G_n^{(s)}$ are symmetric homogeneous 
functions of the 
wave vectors \{ ${\bf q}_1,  \ldots  ,{\bf q}_n $\} with degree zero. They 
are constructed from the fundamental mode coupling functions $\alpha({\bf k}, 
{\bf k}_1)$ and $\beta({\bf k}, {\bf k}_1, {\bf k}_2)$ according to 
the recursion relations ($n \geq 2$, see Goroff et al. (1986),
Jain \& Bertschinger (1994), for a 
derivation):

\begin{mathletters}
\label{ec:recrel}
\begin{eqnarray}
 F_n({\bf q}_1,  \ldots  ,{\bf q}_n) &=& \sum_{m=1}^{n-1}
 { G_m({\bf q}_1,  \ldots  ,{\bf q}_m) \over{(2n+3)(n-1)}} 
 \Bigl[(2n+1) \alpha({\bf k},{\bf k}_1)  F_{n-m}({\bf q}_{m+1},  \ldots  
 ,{\bf q}_n) \nonumber \\ 
&+& 2 \beta({\bf k},{\bf k}_1, {\bf k}_2)  G_{n-m}({\bf q}_{m+1},  
 \ldots  ,{\bf q}_n) \Bigr]
 \label{ec:Fn},
\end{eqnarray}
\begin{eqnarray}
	 G_n({\bf q}_1,  \ldots  ,{\bf q}_n) &=& \sum_{m=1}^{n-1}
 { G_m({\bf q}_1,  \ldots  ,{\bf q}_m) \over{(2n+3)(n-1)}} 
 \Bigl[3 \alpha({\bf k},{\bf k}_1)  F_{n-m}({\bf q}_{m+1},  \ldots  
 ,{\bf q}_n) \nonumber \\
&+& 2n \beta({\bf k},{\bf k}_1, {\bf k}_2)  G_{n-m}({\bf q}_{m+1},  
 \ldots   ,{\bf q}_n) \Bigr]
	\label{ec:Gn},
\end{eqnarray}
\end{mathletters}

\noindent (where ${\bf k}_1 \equiv {\bf q}_1 + \ldots 
+ {\bf q}_m$,  ${\bf k}_2 \equiv 
{\bf q}_{m+1} + \ldots + {\bf q}_n$,  ${\bf k} \equiv {\bf k}_1 +{\bf 
k}_2$, and $F_1= G_1 \equiv 1$) and the symmetrization 
procedure:

\begin{mathletters}
\label{ec:symm}
\begin{equation}
 F_n^{(s)}({\bf q}_1,  \ldots  ,{\bf q}_n) = {1\over{n!}}\sum_{\pi }
 F_n({\bf q}_{\pi (1)},  \ldots  ,{\bf q}_{\pi (n)})
 \label{ec:Fns},
\end{equation}
\begin{equation}
	 G_n^{(s)}({\bf q}_1,  \ldots  ,{\bf q}_n) ={1\over{n!}} \sum_{\pi }
  G_n({\bf q}_{\pi (1)},  \ldots  ,{\bf q}_{\pi (n)})
	\label{ec:Gns},
\end{equation}
\end{mathletters}

\noindent where the sum is taken over all the permutations $\pi $ of the set 
$\{1, \ldots ,n\}$.  For example, for $n=2$ we have:

\begin{mathletters}
\begin{equation}
	 F_2^{(s)}({\bf q}_1,{\bf q}_2) = \frac{5}{7} + \frac{1}{2} 
	 \frac{{\bf q}_1 \cdot {\bf q}_2}{q_1 q_2} \left(\frac{q_1}{q_2} +
	 \frac{q_2}{q_1}\right) + \frac{2}{7} \frac{({\bf q}_1 \cdot {\bf 
	 q}_2)^2}{q_1^2 q_2^2}
	\label{F2},
\end{equation} 
\begin{equation}
	 G_2^{(s)}({\bf q}_1,{\bf q}_2) = \frac{3}{7} + \frac{1}{2} 
	 \frac{{\bf q}_1 \cdot {\bf q}_2}{q_1 q_2} \left(\frac{q_1}{q_2} +
	 \frac{q_2}{q_1}\right) + \frac{4}{7} \frac{({\bf q}_1 \cdot {\bf 
	 q}_2)^2}{q_1^2 q_2^2}
	\label{G2}.
\end{equation} 
\end{mathletters}

The perturbation theory kernels have the following 
properties~(\cite{GGRW86,Wise88}):


\begin{enumerate}
	\item As ${\bf k} = {\bf q}_1 + \dots + {\bf q}_n$ goes to zero, 
	but the individual 
	${\bf q}_i$ do not, $F_n^{(s)} \propto k^2$. This is a consequence of
	momentum conservation in center of mass coordinates. 

	\item  As some of the arguments of $F_n^{(s)}$ or $G_n^{(s)}$ get large 
	but the total sum ${\bf k} = {\bf q}_1 + \dots +{\bf q}_n$ stays fixed, 
	the kernels vanish in inverse square law. That is, for $p \gg q_i$, we 
	have:
	
	\begin{equation}
		F_n^{(s)}({\bf q}_1, \dots ,{\bf q}_{n-2},{\bf p},-{\bf p}) \approx
		G_n^{(s)}({\bf q}_1, \dots ,{\bf q}_{n-2},{\bf p},-{\bf p}) \propto 
		k^2/p^2
		\label{inversesq}. 
	\end{equation}

	\item   If one  of the arguments ${\bf q}_i$ of $F_n^{(s)}$ 
	or $G_n^{(s)}$ goes to 
	zero, there is an infrared divergence of the form ${\bf q}_i /q_i^2$. 
	This comes from the infrared behavior of the mode coupling functions 
	$\alpha({\bf k},{\bf k}_1)$ and $\beta({\bf k}, {\bf k}_1, {\bf k}_2)$. 
	There are no infrared divergences as partial sums of several wavevectors 
	go to zero.
\end{enumerate}

\subsection{Diagrammatic Expansion  of Statistical Quantities}
\label{sec:diagrams}

In this work we focus on the non-linear evolution of two-point cumulants 
of the density field, such as the power spectrum and the 
volume-average two-point 
correlation function, and their 1-point counterpart, the variance. These 
are defined respectively by:

\begin{equation}
	\Big< \tilde{\delta}({\bf k},\tau) \tilde{\delta}
({\bf k}',\tau)\Big>_c = \delta_D({\bf k}+{\bf k}') P(k,\tau)
	\label{powerspectrum},
\end{equation}

\begin{equation}
	\bar \xi(R,\tau) \equiv \int \xi(x,\tau){\bar W}(x)d^3 x =  
\int P(k,\tau) W(k R) d^3 k 
	\label{avgcorrfun},
\end{equation}

\begin{equation}
	\sigma^2(R,\tau) = \int P(k,\tau) \ W^2(k R) \ d^3 k 
	\label{variance}.
\end{equation}

\noindent Here 
$ \xi(x,\tau) = \int P(k,\tau) \ \exp (i {\bf k} \cdot {\bf x}) \ d^3 k$ 
is the two-point correlation function, and 
${\bar W}(x)$ is a window function, with Fourier transform  
$W(kR)$ which we take to be either 
a top-hat (TH) or a Gaussian (G), 

\begin{equation}
	W_{\rm TH} (u)  =   \frac{3}{u^3}\Big[ \sin (u) - u \cos (u) \Big]
	\label{WTH} ,
\end{equation}

\begin{equation}
	W_{\rm G} (u)  =  \exp (-u^2 /2)
	\label{WG}.
\end{equation}

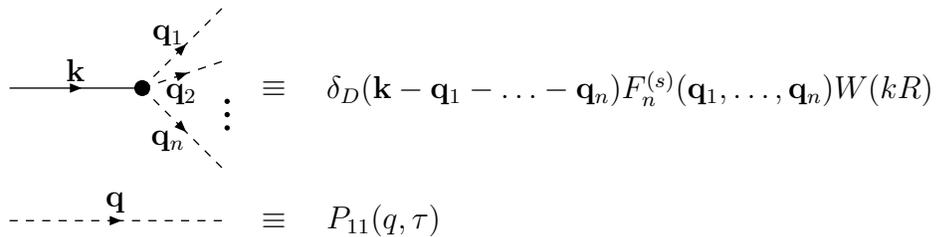
\begin{figure}

\begin{center}\begin{picture}(300,100)(0,-20)

\ArrowLine(0,50)(50,50)
\Vertex(50,50){3}
\DashArrowLine(50,50)(80,80){3}
\DashArrowLine(50,50)(80,60){3}
\DashArrowLine(50,50)(80,20){3}

\Text(25,57)[]{${\bf k}$}
\Text(60,70)[]{${\bf q}_1$}
\Text(65,48)[]{${\bf q}_2$}
\Text(60,30)[]{${\bf q}_n$}

\Vertex(82,45){1}
\Vertex(82,40){1}
\Vertex(82,35){1}

\Text(100,50)[]{$\equiv$}
\Text(120,50)[l]{$ \delta_D({\bf k}- {\bf q}_1- \ldots 
		-{\bf q}_n) F_n^{(s)}({\bf q}_1,\ldots,{\bf q}_n) W(k R)$}

\DashArrowLine(0,0)(80,0){3}
\Text(40,7)[]{${\bf q}$}
\Text(100,0)[]{$\equiv$}
\Text(120,0)[l]{$ P_{11}( q,\tau) $}

\end{picture}\end{center}
\caption{ Diagrammatic rules for vertices and internal lines for density 
field fluctuations smoothed with window $W$ at radius~$R$.}
\label{fig1}

\end{figure}

We are interested in calculating the non-linear evolution of 
these statistical quantities from Gaussian initial conditions
in the weakly non-linear regime, $\sigma(R) \la 1$.  
A systematic framework for  calculating  correlations of cosmological 
fields in perturbation theory has been formulated using
diagrammatic techniques (\cite{GGRW86,Wise88,ScFr96}).  In this 
approach, contributions to $p$-point cumulants 
of the density field come from 
connected diagrams with $p$ external (solid) lines and $r=p-1, p, \dots$ 
internal (dashed) lines.  The perturbation expansion
leads to a collection of diagrams at each order, the leading order being 
tree-diagrams, the next to leading order  1-loop diagrams and so on.  
In each diagram, external lines represent the spectral components 
of the fields we are interested in (e.g., $\delta({\bf k},\tau)$). 
Each internal line is labeled by a wave-vector 
that is integrated over, and represents a linear power spectrum  
$P_{11}(q,\tau)$. Vertices of order $n$ (i.e., where $n$ internal lines 
join) represent a $n^{\rm th}$ order perturbative solution $\delta_n$, 
and momentum conservation is imposed at each 
vertex. Figure~\ref{fig1} shows the factors associated with vertices 
and internal lines. To find the contribution of order $2r$ to the 
{\it $p$-point  spectrum} of the density field proceed as 
follows~(\cite{ScFr96}):


\begin{itemize}
	\item  Draw all distinct connected diagrams containing $p$ vertices 
	(with external lines labeled by ${\bf k}_1 \ldots {\bf k}_p$) 
	joined by $r$ internal lines. Two diagrams are distinct if they cannot 
	be deformed into each other by moving the vertices and lines 
	without cutting any 
	internal lines (sliding lines over other lines is allowed in the 
	rearrangement process). For each of these diagrams:
	
	\begin{enumerate}
		
		\item  Assign a factor of $ \delta_D({\bf k}_i - {\bf q}_1 - 
\ldots -{\bf q}_n) F_n^{(s)}({\bf q}_1,  \ldots ,{\bf q}_n)$  to each vertex 
of order $n$ and external momentum ${\bf k}_i$ ($i=1, \ldots ,p$). 
For the arguments of $F_n^{(s)}$, we  use the convention of assigning 
a positive sign to wave-vectors outgoing from the vertex.
		
\item  Assign a factor of $P_{11}(q_j,\tau)$ to each internal line 
labeled by ${\bf q}_j$. 
		
\item  Integrate over all  ${\bf q}_j$ ($j=1, \ldots ,r$).
		
\item Multiply by the symmetry factor of the graph, which is the 
number of permutations of linear fluctuations ($\delta_1$'s )
that leaves the graph invariant.
		
\item Sum over distinct labelings of external lines, thus
generating $p!/(n_1! \ldots n_{2r-p+1}!)$ diagrams (where $n_i$  
denotes the number of vertices of order $i$).

     \end{enumerate} 
	
\item  Add up the resulting expressions for all these diagrams.
\end{itemize} 

To calculate 1-point cumulants, we have to integrate further over the 
${\bf k}_i$ ($i=1,\dots p$). These 
integrations are trivial because of the presence of  delta functions 
given by rule 1, so we are left only with integrations over the ${\bf 
q}_j$'s. Also, since the $p!/(n_1! \ldots n_{2r-p+1}!)$ diagrams 
generated by rule 5 become equal contributions when the integration over 
external lines is performed, to find the contribution of order $2r$ to the 
{\it $p^{th}$-order 1-point cumulant} of the density field we replace 
rule 5 by the 
following:

\begin{itemize}

\item 5a. Integrate over ${\bf k}_i$ ($i=1,\dots p$) and multiply by 
		 the multinomial weight $p!/(n_1! \ldots n_{2r-p+1}!)$.

\end{itemize}

\noindent Furthermore, when calculating  cumulants of the {\it 
smoothed} density field, we add a window function at each vertex, giving 
instead of rule 1 (see Fig.~\ref{fig1}): 

\begin{itemize}

\item 1a. Assign a factor of $ \delta_D({\bf k}_i - {\bf q}_1 - \ldots 
-{\bf q}_n) F_n^{(s)}({\bf q}_1,  \ldots ,{\bf q}_n) W(k_i R)$  
to each vertex 
of order $n$ and external momentum ${\bf k}_i$ ($i=1, \ldots ,p$).

\end{itemize}


\begin{figure}

\begin{center}\begin{picture}(330,100)(0,0)

\ArrowLine(0,50)(20,50)
\Vertex(20,50){3}
\DashArrowLine(20,50)(70,50){3}
\Vertex(70,50){3}
\ArrowLine(70,50)(90,50)
\Text(45,57)[]{${\bf k}$}
\Text(45,0)[]{$(P_{11})$}

\Text(100,50)[]{+}
\Text(110,50)[]{$\Bigg[$}

\ArrowLine(120,50)(140,50)
\Vertex(140,50){3}
\DashCurve{(140,50)(165,60)(190,50)}{3}
\DashCurve{(140,50)(165,40)(190,50)}{3}
\DashArrowLine(164,60)(166,60){3}
\DashArrowLine(164,40)(166,40){3}
\Vertex(190,50){3}
\ArrowLine(190,50)(210,50)
\Text(165,67)[]{${\bf k}-{\bf q}$}
\Text(165,33)[]{${\bf q}$}
\Text(165,0)[]{$(P_{22})$}

\Text(220,50)[]{+}

\ArrowLine(230,50)(250,50)
\Vertex(250,50){3}
\DashArrowArcn(250,60)(10,0,180){3}
\DashArrowArcn(250,60)(10,180,360){3}
\DashArrowLine(250,50)(300,50){3}
\Vertex(300,50){3}
\ArrowLine(300,50)(320,50)
\Text(250,77)[]{${\bf q}$}
\Text(275,57)[]{${\bf k}$}
\Text(275,0)[]{$(P_{13})$}
\Text(330,50)[]{$\Bigg]$}

\end{picture}\end{center}
\caption{ Diagrams  for the power spectrum
 up to one loop. See Eqs.~(\protect \ref{P1lamp}) 
for diagram amplitudes.}
\label{fig2}

\end{figure}
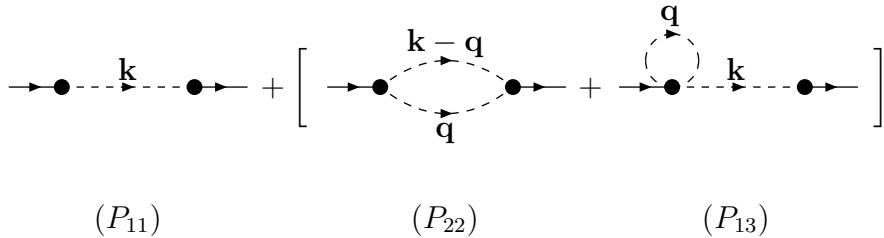

According to the above diagrammatic rules, we can write the loop 
expansion for the power spectrum up to one-loop corrections as 

\begin{equation}
	P(k,\tau) = P^{(0)}(k,\tau) + P^{(1)}(k,\tau) + \ldots 
	\label{Ploopexp},
\end{equation}

\noindent where the superscript $(n)$ denotes an $n$-loop contribution, 
the tree-level ($0$-loop) contribution is just the linear spectrum,

\begin{equation}
	P^{(0)}(k,\tau) =  P_{11}(k,\tau)  
	\label{P^(0)},
\end{equation}

\noindent with $a^2(\tau) \langle \delta_1({\bf k}) \delta_1({\bf k}')\rangle_c 
= \delta_D({\bf k}+{\bf k}') P_{11}(k,\tau)$, 
and the 1-loop contribution consists of two terms,

\begin{equation}
	P^{(1)}(k,\tau) = P_{22}(k,\tau) + P_{13}(k,\tau) 
	\label{P^(1)},
\end{equation}

\noindent with (see Fig.~\ref{fig2}):

\begin{mathletters}
\label{P1lamp}
\begin{eqnarray}
	 P_{22}(k,\tau) & \equiv  &  2 \int [F_2^{(s)}({\bf k}-{\bf q},{\bf q}) ]^2
	 P_{11}(|{\bf k}-{\bf q}|,\tau)  P_{11}(q,\tau)  d^3q
	\label{P22},  \\
	 P_{13}(k,\tau)  & \equiv &   6  \int 
	 F_3^{(s)}({\bf k},{\bf q},-{\bf q}) P_{11}(k,\tau)	
	 P_{11}(q,\tau)  d^3q
	\label{P13}.
\end{eqnarray}
\end{mathletters}

\noindent Here $P_{ij}$ denotes the 
amplitude given by the above rules
for a connected diagram representing the contribution from  
$\langle \delta_i  \delta_j \rangle_c$ to the power spectrum. We have 
assumed Gaussian initial conditions, for which  
$P_{ij}$ vanishes if $i+j$ is odd.

For the smoothed variance and average two-point correlation function we write

\begin{mathletters}
\label{sigxiexp}
\begin{equation}
	\sigma^2(R)  = 
	\sigma_\ell^2(R) \Big( 1 + s^{(1)} \ \sigma_\ell^2(R)
	+ \ldots \Big)
	\label{siexp},
\end{equation} 
\begin{equation}
	\bar \xi(R) = 
	\bar \xi_\ell(R) \Big( 1 + x^{(1)} \ \bar \xi_\ell(R)
	+ \ldots \Big)
	\label{xiexp},
\end{equation} 
\end{mathletters}

\noindent where $\sigma_\ell^2(R)$ and $\bar \xi_\ell(R)$ denote the 
variance and average two-point correlation function in linear theory 
(given by (\ref{variance}) and (\ref{avgcorrfun}) with $P=P_{11}$); 
the dimensionless 1-loop amplitudes are 

\begin{mathletters}
\label{s1x1}
\begin{eqnarray}
	 s^{(1)}(R) & \equiv & \frac{1}{\sigma_\ell^4(R)}
	 \int P^{(1)}(k,\tau) \ W^2(k R) \ d^3 k
	\label{s1}, \\
	 x^{(1)}(R) & \equiv & \frac{1}{\bar \xi^2_\ell(R)}
	 \int P^{(1)}(k,\tau) \ W(k R) \ d^3 k
	\label{x1}. 
\end{eqnarray} 
\end{mathletters}

A convenient property of these dimensionless amplitudes is the following: 
if one defines the correlation length $R^{(\ell)}_0$ in linear theory as 
the scale where the smoothed linear variance is unity,  
$\sigma^2_\ell(R^{(\ell)}_0)=1$, then    
$s^{(1)}(R^{(\ell)}_0) = \sigma^2(R^{(\ell)}_0)-1$ is just the 
non-linear correction to the variance at this scale.


\section{Results} 
\label{sec:results}

\subsection{One-Loop Power Spectrum} 
\label{sec:1lps}

We consider a linear (tree-level) power spectrum $P_{11}(k,\tau)$ given by a
truncated power-law,

\begin{equation}
 P_{11}(k,\tau) \equiv \left\{ \begin{array}{ll} 
                               A \ a^2(\tau) \ k^n & \mbox{if $\epsilon \leq 
                               k \leq k_c$}, \\
                               0 & \mbox{otherwise},
                               \end{array}
                       \right. 
\label{P11}
\end{equation}

\noindent where $A$ is a normalization constant, and the infrared and 
ultraviolet cutoffs $\epsilon$ and 
$k_c$ are imposed in order to regularize the required radial
integrations~(\cite{ScFr96}). In a cosmological N-body simulation, they would 
correspond roughly to the inverse comoving box size and lattice spacing (or 
interparticle separation) respectively. 
In the absence of the cutoffs, the spectrum 
(\ref{P11}) would be scale-free. 
We write the one-loop power spectrum contributions as

\begin{equation}
 	P_{ij}(k,\tau) \equiv \pi A^2 a^4(\tau) \ k_c^{2n+3} \ 
p_{ij}(n;x,\Lambda)
 	\label{P2def},
\end{equation} 

\noindent where the dimensionless 1-loop spectrum $p^{(1)}=p_{22}+p_{13}$, and  
we have introduced dimensionless wavenumber variables, 

\begin{equation}
 x \equiv k/k_c, \ \ \ \ \ \ \ \ \ \ \Lambda \equiv k_c / \epsilon 
\label{dimenq}.
\end{equation} 
 
 \noindent From Eq.~(\ref{P22})~ we obtain (defining $t \equiv q/k_c$):

 \begin{equation}
 	p_{22}(n;x,\Lambda) \equiv  \frac{x^4}{49} \int_{1/ \Lambda}^1
 	dt \ t^n \int_{\lambda_{min}(x,t)}^{\lambda_{max}(x,t,\Lambda)}
 	 d \lambda \ 
 	(x^2+t^2 -2 x t \lambda)^{n/2-2} \ (3 t - 7 x \lambda - 10 t \lambda^2)^2
 	\label{p22},
 \end{equation} 

\noindent when  $x \leq 2$; otherwise $p_{22}(n;x,\Lambda) =0$.
The constraint on the angular integration 
variable $\lambda \equiv
({\bf k} \cdot {\bf q})/ (k q)$ comes from the cutoff dependence of
$ P_{11}(|{\bf k}-{\bf q}|,\tau)$, which gives:

 \begin{mathletters}
\label{lambdamm}
 \begin{equation}
 	  \lambda_{min}(x,t) \equiv {\rm Max} \left\{-1, \frac{x^2+t^2 -1}{2 x t}
 	  \right\}
 	\label{lambdamin},
 \end{equation} 

 \begin{equation}
 	\lambda_{max}(x,t,\Lambda) \equiv  {\rm Min} 
 	\left\{1, \frac{x^2+t^2 -\Lambda^{-2}}{2 x t} \right\}
 	\label{lambdamax},
 \end{equation} 
 \end{mathletters}

\noindent Care must be taken when dealing with the limits imposed by 
Eqs.~(\ref{lambdamin})~and~(\ref{lambdamax}), especially in the cases
$n=-1,-2$. On the other hand, integration over angular variables in
Eq.~(\ref{P13}) is straightforward, and we obtain:
 
 \begin{eqnarray}
 	 p_{13}(n;x,\Lambda) & \equiv  & x^n \int_{1/ \Lambda}^1 dt \ t^{n+2}
 \Bigg[ \frac{(6 x^6  - 79 x^4  t^2  + 50 x^2  t^4  - 21 t^6 )}{63 x^2 t^4} +
 	 \frac{(t^2-x^2)^3 \ (7 t^2 + 2 x^2 )}{42 x^3 t^5} \nonumber \\
 	 & & \times \ln \frac{|x+t|}{|x-t|} \Bigg]
 	\label{p13},
 \end{eqnarray}

\noindent for $ \Lambda^{-1} \leq x \leq 1$; otherwise $p_{13}(n;x,\Lambda)
 = 0$.

In the following subsections 
we give results for $p_{13}$ and  $p_{22}$ in the limit
 $\Lambda =k_c/\epsilon \rightarrow \infty$ up to 
terms of order $\Lambda^{0}$, for spectral indices $n=1,0,-1,-2$. We 
also give their asymptotic behavior at large scales. From the properties of 
the perturbation theory kernels given in Sec.~\ref{sec:recrel} and 
Eqs.~(\ref{P1lamp}), one would 
naively expect that as $x \rightarrow 0$: 

\begin{equation}
	 p_{22}(n;x,\Lambda) \propto x^4 \sim (k/k_c)^4 ,  \ \ \ \ \ \ \ \ \ \
	 p_{13}(n;x,\Lambda)   \propto   x^{n+2} \sim (k/k_c)^{n+2}
	\label{p1lasymp}.
\end{equation}

\noindent Although this is certainly correct for $ p_{22}$ 
when $x \ll \Lambda^{-1}$ (i.e., $k \ll \epsilon$), 
the expressions below correspond to $ p_{22}$ and $ p_{13}$ {\it after} the 
limit $\Lambda \rightarrow \infty$ has been taken and therefore exhibit 
a different kind of asymptotic behavior (corresponding to 
$0 \leftarrow \epsilon < k \ll 
k_c$). This deviation from the expected scaling becomes more pronounced as $n$ 
decreases, since infrared effects ($\epsilon \rightarrow 0$) become more 
important with the increase of large-scale power.  As we will see, 
Eq.~(\ref{p1lasymp}) is obeyed by $p_{22}$ only when $n \geq 1$ and by $p_{13}$ 
when $n \geq 0$.

Our 
results below are equivalent to those of Makino et al. (1992)
 in the case of $p_{13}$, 
but they differ for $p_{22}$ in two respects. First, their expressions 
for $p_{22}$ are only valid for $x < 1$; they did not seem to consider 
that $p_{22}$ is non-vanishing up to $x=2$, and the region $1 \leq x \leq 2$ 
requires a separate integration when $n$ is odd. In addition, for $n$ even, 
their series expansions for the 
dilogarithms do not converge in this region. 
The second difference comes from the fact that 
for $n=-1,-2$, $p_{22}$ develops a divergence at $x=1$, and the 
expressions for $p_{22}$ must be modified in a region of radius 
$\Lambda^{-1}$ about $x=1$. This singularity is integrable, giving a finite 
contribution to the 1-loop correction to the variance. Finally, some 
of their resulting expressions for $p^{(1)}$ contain typographic errors. 
We have checked our expressions by analytically integrating them 
and verifying that they correctly reproduce 
the unsmoothed 1-loop coefficients $s^{(1)}$ given by
Scoccimarro \& Frieman (1996). For comparison, we also plot (but do 
not give explicit expressions for) the 1-loop power spectrum in the 
Zel'dovich approximation. 


\subsubsection{$n=1$} 
\label{sec:1lpsn1}

\begin{figure}[h]
\centering
\centerline{\epsfxsize=10. truecm \epsfysize=9. truecm \epsfbox{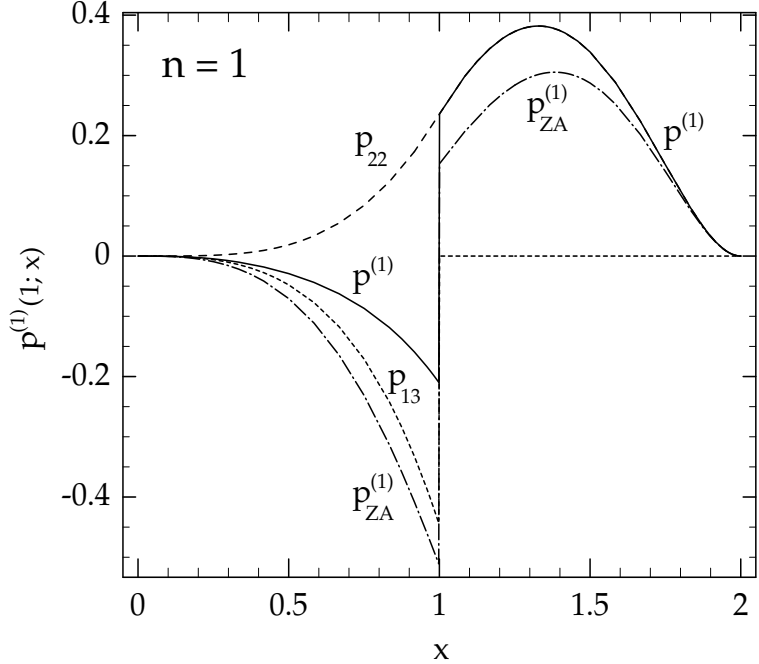}}

\caption{One-loop corrections to the power spectrum for $n=1$. 
The solid line shows $p^{(1)}(1;x)$ as a function of $x \equiv k / k_c$. 
Dashed lines correspond to the contribution of individual diagrams, 
$p_{22}(1;x)$ 
(long-dashed) and $p_{13}(1;x)$ (short-dashed), whereas the dot-dashed 
line shows $p^{(1)}(1;x)$ in the Zel'dovich approximation.}
\label{p1ln1}
\end{figure}

For $n=1$, the dimensionless 1-loop power spectrum contributions are 

\begin{eqnarray}
	 p_{13}(1;x)  & = &   -\frac{1}{21 x}  + 
	 \frac{52 x}{315} - \frac{181 x^3}{315}  -
 \frac{2 x^5}{21} + \ln \left( 
\frac{1+x}{1-x} \right) \frac{(5 - 19 x^2 + 25 x^4 - 5 x^6 + 10 x^8)}{210 x^2} 
	 \nonumber \\ 
& &+ \frac{8 x^5}{105}  \ln \Big(\frac{1-x^2}{x^2} \Big) 
	\label{f1},  
\end{eqnarray}

\begin{equation}
 p_{22}(1;x) = \frac{18 x^4}{49}  - 	\frac{13 x^5}{98} - 
 \frac{20 x^6}{1029} +\frac{x^7}{49},  \ \ \ \ \ \ {\rm if} \ x \leq 1, 
\label{g1a}
\end{equation}

\begin{equation}
p_{22}(1;x) =  -\frac{320 }{1029 x} + \frac{16 x }{49} + 
 \frac{2 x^3}{3} -\frac{18 x^4}{49}  -
  \frac{5 x^5}{42} + \frac{20 x^6}{1029}  + 
  \frac{x^7}{49}, \ \ \ \ \   \ {\rm if} \ 1 \leq x \leq 2,
\label{g1b}
\end{equation} 

\noindent and the 1-loop reduced power spectrum is then (see Fig. {\protect 
\ref{p1ln1}}):

\begin{eqnarray}
	 p^{(1)}(n=1;x=k/k_c \leq 1) &=&  -\frac{1}{21 x}  + 
	 \frac{52 x}{315} - \frac{181 x^3}{315} + \frac{18 x^4}{49}  -
 \frac{67 x^5}{294} -  \frac{20 x^6}{1029} +\frac{x^7}{49}  \nonumber \\ 
& & + \ln \left( 
\frac{1+x}{1-x} \right)
 \frac{(5 - 19 x^2 + 25 x^4 - 5 x^6 + 10 x^8)}{210 x^2} 
	 \nonumber \\
& & + \frac{8 x^5}{105} 
	 \ln \left(\frac{1-x^2}{x^2} \right),
	\label{p1}  
\end{eqnarray}

\noindent with $p^{(1)}(x) = p_{22}(x)$ when $1 \leq x \leq 2$.
 For $x \ll 1$ we get:

\begin{eqnarray}
	p_{13}(1;x) & \approx & p^{(1)}(1;x)
	 \approx - \frac{122}{315} x^3 + 
	{\cal O}(x^5)
	\label{f1x0},  \\
	p_{22}(1;x) & \approx &   \frac{18}{49} x^4 + 
	{\cal O}(x^5)
	\label{g1x0}.
\end{eqnarray}

\subsubsection{$n=0$} 
\label{sec:1lpsn0}

For $n=0$, the 1-loop power spectrum contributions are
 
\begin{eqnarray}
p_{13}(0;x) &=&  -  \frac{1}{18 x^2} +  \frac{157}{756} - \frac{269 x^2}{252} 
- \frac{ \pi^2 x^3}{84} - \frac{ x^4}{21} 
+ \ln \left(\frac{1+x}{1-x} \right) \nonumber \\ & &\times  
\frac{(x^2-1)}{504 x^3} (-14 + 43 x^2 - 
47 x^4 + 12 x^6)  + \frac{ x^3}{42} [ {\rm Li}_2(x) - {\rm Li}_2 (-x) ] 
	\label{f0},  
\end{eqnarray}

\begin{eqnarray}
p_{22}(0;x\leq 2) & = &  \frac{25}{98}+ \frac{25 x}{196}-\frac{10 x^2}{147}-
\frac{ 5 x^3}{392} + \frac{ 29 \pi^2 x^3}{294} + \frac{3 x^4}{49} 
- \frac{65 x^5}{1176} + \frac{x^7}{196} + \ln |x-1| \nonumber \\
& & \times  \frac{(25 - 15 x^2 - 16 x^4 + 6 x^6 + 29 x^4 \ln x)}{98 x} 
	 - \frac{29x^3}{98} \Big[ \frac{1}{2} \ln^2 (x) + 
	{\rm Re} [ {\rm Li}_2 (x) ] \nonumber \\
& & + 	{\rm Li}_2 \left( \frac{x-1}{x} \right) \Big] 
\label{g0}, 
\end{eqnarray}

\noindent where  
${\rm Li}_2$ denotes the dilogarithm, defined by~(\cite{Lewin81})

\begin{equation}
{\rm Li}_2(x) \equiv - \int_0^x dz \ \frac{\ln (1-z)}{z}
\label{dilog},
\end{equation}

\noindent which has the series expansion for small argument ($x \leq 1$): 

\begin{equation}
{\rm Li_2}(x) = \sum_{m=1}^{\infty}  \frac{x^m}{m^2}
\label{dilogseries}.
\end{equation}

\begin{figure}[t]
\centering
\centerline{\epsfxsize=10. truecm \epsfysize=9. truecm \epsfbox{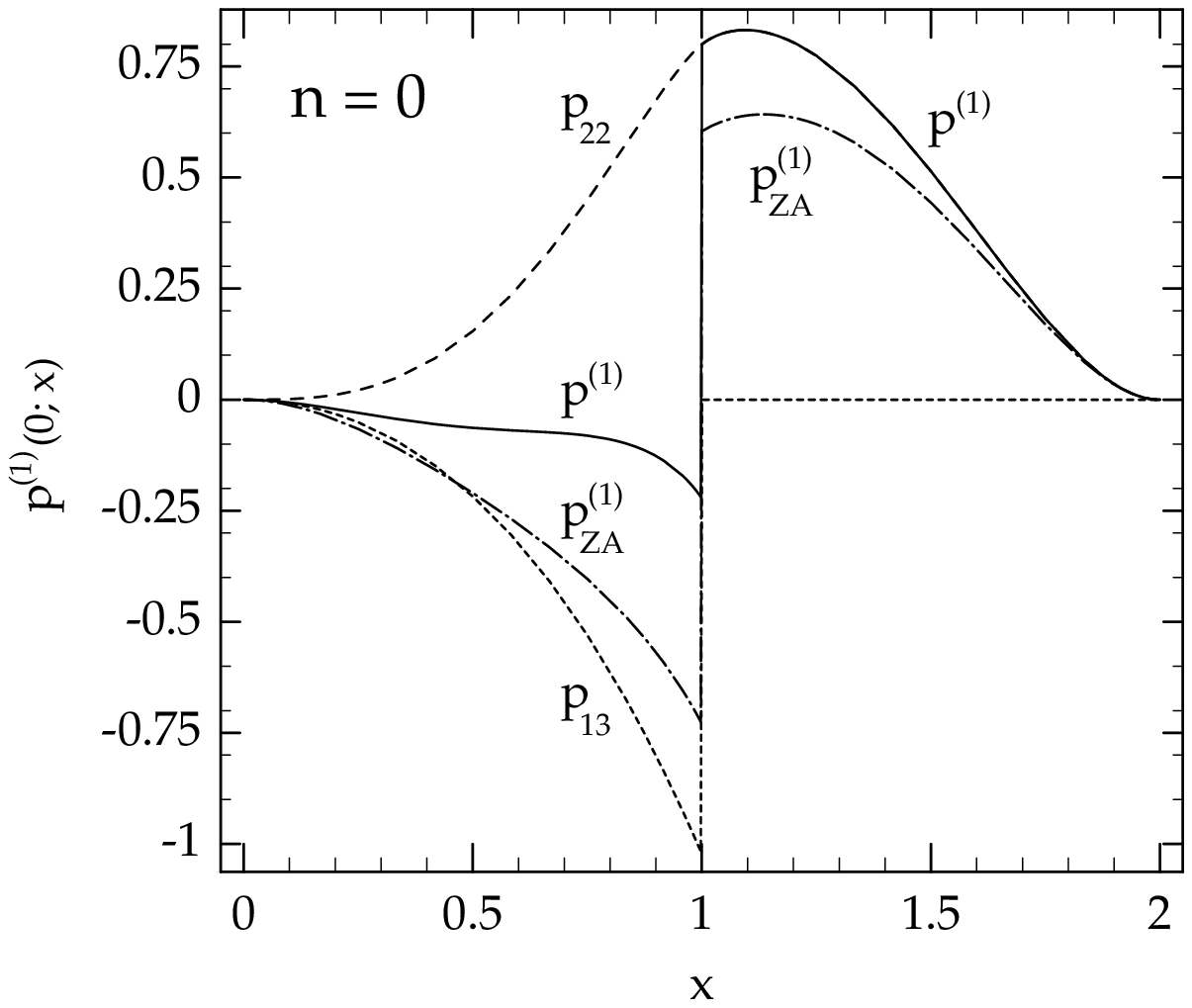}}

\caption{Same as Fig.~{\protect \ref{p1ln1}} but for $n=0$.}
\label{p1ln0}
\end{figure}

\noindent The resulting 1-loop power spectrum is (Fig. {\protect \ref{p1ln0}})

\begin{eqnarray}
	p^{(1)}(0;x\leq 1) & = &   - \frac{1}{18 x^2} +  \frac{2449}{5292} +
	 \frac{25 x}{196} - \frac{2003 x^2}{1764}  -
\frac{ 5 x^3}{392} + \frac{ 17 \pi^2 x^3}{196} + \frac{2 x^4}{147} 
- \frac{65 x^5}{1176} + \frac{x^7}{196} \nonumber   \\
& & + \ln (1+x) \frac{(x^2-1)}{504 x^3} (-14 + 43 x^2  
	  - 47 x^4 + 12 x^6) - \frac{29x^3}{196} \ln^2 (x) \nonumber \\ & & + 
\frac{(-98 +  1299 x^2 - 1170 x^4 - 163 x^6 + 132 x^8 + 
1044 x^6 \ln x)}{3528 x^3} \ln (1-x) \nonumber   \\ & &
 - 	\frac{40 x^3}{147} {\rm  Li}_2 (x) 
- \frac{ x^3}{42} {\rm Li}_2 (-x)  - 	\frac{29 x^3}{98} {\rm Li}_2 \left( 
\frac{x-1}{x} \right),   
	\label{p0}  
\end{eqnarray}

\noindent and $p^{(1)}(x) = p_{22}(x)$ when $1 \leq x \leq 2$.

For $x \ll 1$ we have:

\begin{eqnarray}
	p_{13}(0;x) & \approx & p^{(1)}(0;x) \approx - \frac{244}{315} x^2 + 
	{\cal O}(x^3)
	\label{f0x0},  \\
	p_{22}(0;x) & \approx &   \frac{29 \pi^2}{196} x^3 + 
	{\cal O}(x^4)
	\label{g0x0}.
\end{eqnarray}

\subsubsection{$n=-1$} 
\label{sec:1lpsnm1}

For $n \leq -1$, the 1-loop contributions are infrared-divergent and 
therefore depend on the infrared cutoff through $\Lambda = k_c/\epsilon$,  

\begin{eqnarray}
p_{13}(-1;x,\Lambda)  & = &- \frac{4 x}{3} \ln \Lambda  -\frac{1}{15 x^3}  + 
	  \frac{88 }{315 x} -  \frac{11 x}{189} - \frac{2 x^3}{63}  
	   + \ln \left( \frac{1+x}{1-x} \right) \nonumber \\ & & \times 
	   \frac{(21 - 95 x^2 + 225 x^4 + 15 x^6 + 10 x^8)}{630 x^4} + 
\frac{88 x}{315}  \ln \left(\frac{1-x^2}{x^2} \right) 
	\label{fm1},  
\end{eqnarray}

\begin{equation}
 p_{22}(-1;x,\Lambda) = \frac{4 x}{3} \ln \Lambda +
 \frac{80 x}{147}  + 	\frac{2 x^2}{3} + 
 \frac{ x^3}{3} +\frac{44 x^4}{441} +\frac{2 x^5}{49}
 +\frac{ x^7}{441} + \frac{2 x}{3} \ln [(1-x) x^2],  
  \ \ \ \ \  \ x \leq 1-\Lambda^{-1}, 
\label{gm1a}
\end{equation} 

\begin{eqnarray}
 p_{22}(-1;x,\Lambda) &=& \frac{2 x}{3} \ln \Lambda +
 \frac{160 }{147 x}  + 
 \frac{ x^3}{3}  +\frac{2 x^5}{49}
 +\frac{ x^7}{441} + \frac{2 x}{3} \ln (x) - \frac{(x-1)^2}{147 x}
 (50 + 100 x \nonumber \\ 
&  &   + 105 x^2 + 12 x^3  + 6 x^4 ) \ \Lambda - \frac{(x-1)^3}{882 x}
 (15 + 45 x + 60 x^2  + 60 x^3  + 12 x^4  \nonumber \\
& &  + 4 x^5 ) \Lambda^3,  
  \ \ \ \ \  1-\Lambda^{-1} \leq x \leq 1+\Lambda^{-1} 
\label{gm1akc}
\end{eqnarray} 
 
\begin{equation}
 p_{22}(-1;x)  =  \frac{320 }{147 x}  - \frac{80 x}{147}  - 	
 \frac{2 x^2}{3} +  \frac{ x^3}{3} - \frac{44 x^4}{441} +\frac{2 x^5}{49}
 +\frac{ x^7}{441} - \frac{2 x}{3} \ln (x-1), 
 \ \ \ \ \   \ 1+\Lambda^{-1} \leq x \leq 2,
\label{gm1b}
\end{equation}

\begin{figure}[t]
\centering
\centerline{\epsfxsize=10. truecm \epsfysize=9. truecm \epsfbox{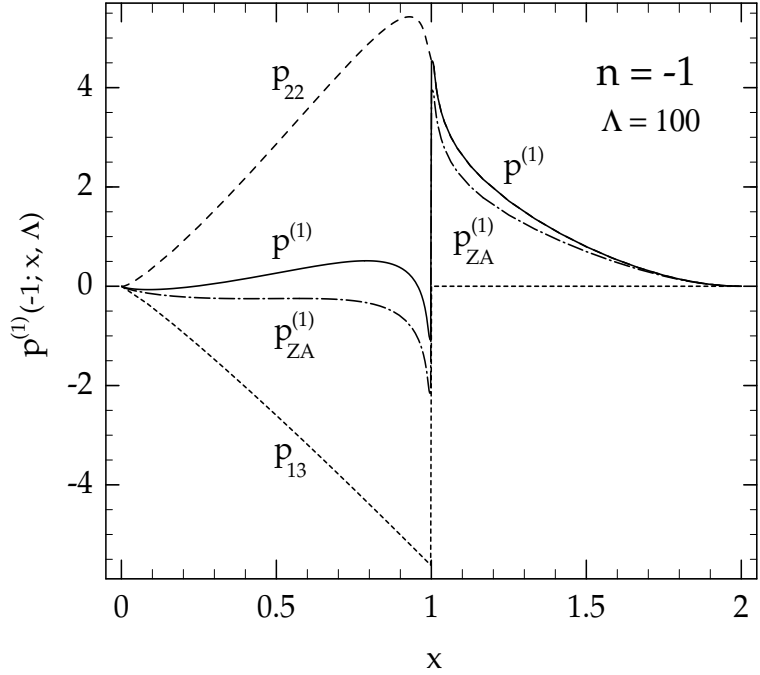}}

\caption{Same as Fig.~{\protect \ref{p1ln1}} but for $n=-1$ and $\Lambda 
\equiv k_c / \epsilon = 100$.}
\label{p1lnm1}
\end{figure}

\noindent The 1-loop correction is then

\begin{eqnarray}
	p^{(1)}(-1;x)   & = &   -\frac{1}{15 x^3}  + 
	  \frac{88 }{315 x} +  \frac{643 x}{1323} + \frac{2 x^2}{3}  + 
 \frac{19 x^3}{63} +\frac{44 x^4}{441} +\frac{2 x^5}{49}
 +\frac{ x^7}{441} \nonumber \\
& &  + \frac{(21 - 95 x^2 + 225 x^4 + 15 x^6 + 10 x^8)}{630 x^4} \ln 
\left( \frac{1+x}{1-x} \right) + \frac{298 x}{315}  \ln (1-x)  
\nonumber \\ & &    + \frac{244 x}{315}  \ln (x)
	   + \frac{88 x}{315}  \ln (1+x), \ \ \   \ x \leq 1-\Lambda^{-1}
	\label{pm1},  
\end{eqnarray}

\noindent whereas when $1 \leq x \leq 2$ we have $p^{(1)}(-1;x) = p_{22}(-1;x)$.
For $x \ll 1$ we get:


\begin{eqnarray}
	p_{13}(-1;x,\Lambda)  & \approx &  - \frac{4 x}{3} \ln \Lambda +
	\frac{128}{225} x - \frac{176 x}{315} \ln (x) + 
	{\cal O}(x^3)
	\label{fm1x0},  \\
	p_{22}(-1;x,\Lambda)  & \approx &   \frac{4 x}{3} \ln \Lambda + 
	\frac{80}{147} x + \frac{4 x}{3} \ln (x) +
	{\cal O}(x^4)
	\label{gm1x0}, \\ 
	p^{(1)}(-1;x) & \approx &  \frac{12272}{11025} x + \frac{244 x}{315} \ln (x) +
	{\cal O}(x^3)
	\label{pm1x0}.
\end{eqnarray}
 
\noindent In this limit, the divergences cancel; note, 
however, that the $\Lambda 
\rightarrow \infty$ divergence in $p_{22}$ in the region $1-\Lambda^{-1} 
\leq x \leq 1 + \Lambda^{-1}$ is uncancelled, although the `size' of 
the divergent region shrinks to zero. 
 

\subsubsection{$n=-2$} 
\label{sec:1lpsnm2}

For $n=-2$, the 1-loop terms are 

\begin{eqnarray}
	p_{13}(-2;x,\Lambda)  & = & -\frac{4}{3} \Lambda   - 
	 \frac{1}{12 x^4} + \frac{107}{252 x^2} + \frac{5 \pi^2}{28 x} 
	 +\frac{82}{63} -  \frac{ x^2}{42}   + \ln \left( 
\frac{1+x}{1-x} \right) \frac{(x^2-1)}{168 x^5} \nonumber \\ & & \times 
(-7 + 31 x^2 + 4 x^4 + 2 x^6) + \frac{5}{14 x} 
 [ {\rm Li}_2 (-x) - {\rm Li}_2 (x) ] 
	\label{fm2},  
\end{eqnarray}


\begin{eqnarray}
	p_{22}(-2;x,\Lambda)  & = & \frac{2}{3}  \Lambda  
	 +  \frac{75}{1568 x} + \frac{25 \pi^2}{196 x} 
	 + \frac{15 x}{392}   +
\frac{ 29 x^3}{392} +   \frac{3 x^5}{196} + \frac{x^7}{784} + 
\frac{\Lambda^2}{196 x} (-50 + 82  x^2  \nonumber \\
& &   - 29 x^4  - 3 x^6 - 30 x^2 \ln x)  +  \frac{\Lambda^4}{1568 x} 
	(25 - 60  x^2 + 63 x^4 - 24 x^6 - 4 x^8  \nonumber   \\
	 & & + 44 x^4 \ln x),  
	\ \ \ \ \ 1-\Lambda^{-1} \leq x \leq 1+\Lambda^{-1} \nonumber \\
	\label{gm2kc}  
\end{eqnarray}

\noindent whereas when  $x \leq 1-\Lambda^{-1}$ or  $ x 
\geq 1+\Lambda^{-1}$ we have:

\begin{eqnarray}
	p_{22}(-2;x,\Lambda)  & = & \frac{4}{3} \theta (1-x) \Lambda  
	 + \frac{1 }{2 (x-1)}  + \frac{25 \pi^2}{98 x} - \frac{205}{392}
	  - \frac{309 x}{784} - \frac{11 x^2}{392}  +
\frac{ 53 x^3}{1568} +   \frac{3 x^5}{196} \nonumber \\ & & + \frac{x^7}{784}
- \frac{75}{196 x} \ln^2 (x) 
+ \ln |x-1| \frac{(-1001 + 60 x^2 - 11 x^4 + 300  \ln x)}{392 x} \nonumber \\ 
& & - \frac{75}{98 x} \left[{\rm Re} [ {\rm Li}_2 (x) ] + 
{\rm Li}_2 \left( 
\frac{x-1}{x} \right)\right],  
	\label{gm2}  
\end{eqnarray}

\noindent where $\theta(x)$ is the step function. This gives:


\begin{eqnarray}
p^{(1)}(-2;x)   & = &   - 	 \frac{1}{12 x^4} + \frac{107}{252 x^2} + 
	 \frac{1 }{2 (x-1)} + \frac{85 \pi^2}{196 x} +  \frac{2747}{3528}  
	  - \frac{309 x}{784} - \frac{61 x^2}{1176}  +
\frac{ 53 x^3}{1568} +   \frac{3 x^5}{196} \nonumber \\ & &
+ \frac{x^7}{784}
- \frac{75}{196 x} \ln^2 (x) 
 + \ln (1-x)  \frac{(-1001 + 60 x^2 - 11 x^4 + 300  \ln x)}{392 x} 
\nonumber \\ & & + \ln \left( 
\frac{1+x}{1-x} \right) \frac{(x^2-1)}{168 x^5} (-7 + 31 x^2 + 4 x^4 + 2 x^6)
 + \frac{5}{14 x} {\rm  Li}_2 (-x) \nonumber   \\
	 &  &  - \frac{55}{49 x} {\rm  Li}_2 (x)  
	 - 	\frac {75}{98 x} {\rm Li}_2 \Big( 
\frac{x-1}{x} \Big),  \ \ \ \ \  \ x \leq 1-\Lambda^{-1},
	\label{pm2}  
\end{eqnarray}

\noindent otherwise $ p^{(1)}(-2;x) = p_{22}(-2;x)$ when $1 + \Lambda^{-1}
 \leq x \leq 2$.For $x \ll 1$ we have:

\begin{eqnarray}
	p_{13}(-2;x,\Lambda) & \approx & - \frac{4}{3} \Lambda +
	\frac{5 \pi^2}{28 x} + {\cal O}(x^0)
	\label{fm2x0},  \\
	p_{22}(-2;x,\Lambda) & \approx &  \frac{4}{3} \Lambda + 
	\frac{75 \pi^2}{196 x}  + {\cal O}(x^0)
	\label{gm2x0}, \\
	p^{(1)}(-2;x) & \approx &   \frac{55 \pi^2}{98 x}  + 
	{\cal O}(x^0)
	\label{pm2x0}.
\end{eqnarray}

\begin{figure}[h!]
\centering
\centerline{\epsfxsize=10. truecm \epsfysize=9. truecm \epsfbox{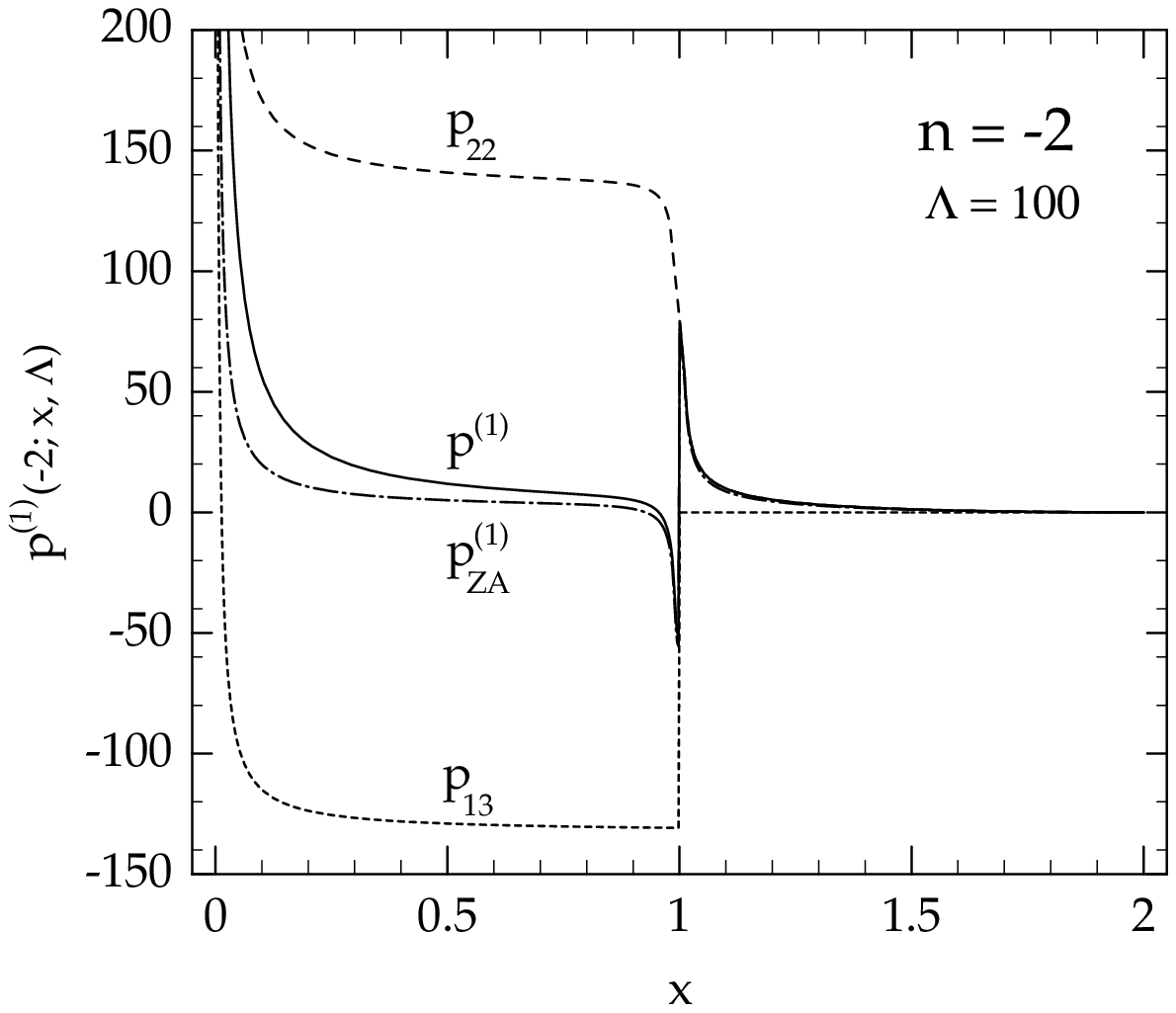}}

\caption{Same as Fig.~{\protect \ref{p1ln1}} but for $n=-2$ and $\Lambda 
\equiv k_c / \epsilon = 100$. For reasons 
of clarity, we do not show the behavior for $x \leq 2 \Lambda^{-1}$ 
($k \leq 2\epsilon$), where 
$p_{22}$ turns over to the scaling given in Eq.~({\protect 
\ref{p1lasymp}}). Similarly, $p_{13} =0$ when $x \leq  \Lambda^{-1}$.} 
\label{p1lnm2}
\end{figure}
 

\subsection{One-Loop Smoothed Variance and Average Two-point Correlation Function} 
\label{sec:1lsv}

Using the results of the previous section for the power spectrum, 
we can calculate the 1-loop 
corrections to the variance and average two-point function, 
Eqs.~(\ref{variance}, \ref{avgcorrfun}). We focus on the corresponding 
1-loop amplitudes $s^{(1)}$ and $x^{(1)}$ defined in Eqs.~(\ref{sigxiexp}, 
\ref{s1x1}). From (\ref{s1x1}), (\ref{P11}), and (\ref{P2def}) we can write:

\begin{mathletters}
\label{s1x1ints}
\begin{equation}
	 s^{(1)}(n;k_c R)  \equiv  \frac{(k_c R)^{2n+3}}{4 
	 [\Delta I_\sigma(n,k_c R,\Lambda)]^2 }
	 \int_0^{2 k_c R} p^{(1)}(n;u (k_c R)^{-1},\Lambda) \ W^2(u) \ u^2 du
	\label{s1int},
\end{equation} 
\begin{equation} 
	 x^{(1)}(n;k_c R) \equiv  \frac{(k_c R)^{2n+3}}{4 
	 [\Delta I_\xi(n,k_c R,\Lambda)]^2 }
	 \int_0^{2 k_c R} p^{(1)}(n;u (k_c R)^{-1},\Lambda) \ W(u) \ u^2 du
	\label{x1int}, 
\end{equation} 
\end{mathletters}

\noindent where $\Delta I_\sigma(n,k_c R,\Lambda) \equiv 
I_\sigma(n,k_c R) - I_\sigma(n,\epsilon R)$ and similarly for
$\Delta I_\xi$, with

\begin{mathletters}
\label{Isigxi}
\begin{eqnarray}
	I_\sigma (n, z) & \equiv  & \int_0^z u^{n+2} \  W^2(u) \ du
	\label{Isigma1st}, \\ 
	I_\xi (n, z) & \equiv & \int_0^z u^{n+2} \  W(u) \ du
	\label{Ixi1st}. 
\end{eqnarray}
\end{mathletters}

\noindent In Appendix~\ref{app:IsigIxi} we present analytic 
results for $I_{\sigma}$ and $I_{\xi}$ with top-hat and Gaussian smoothing.

\begin{figure}[t!]
\centering
\centerline{\epsfxsize=10. truecm \epsfysize=9. truecm \epsfbox{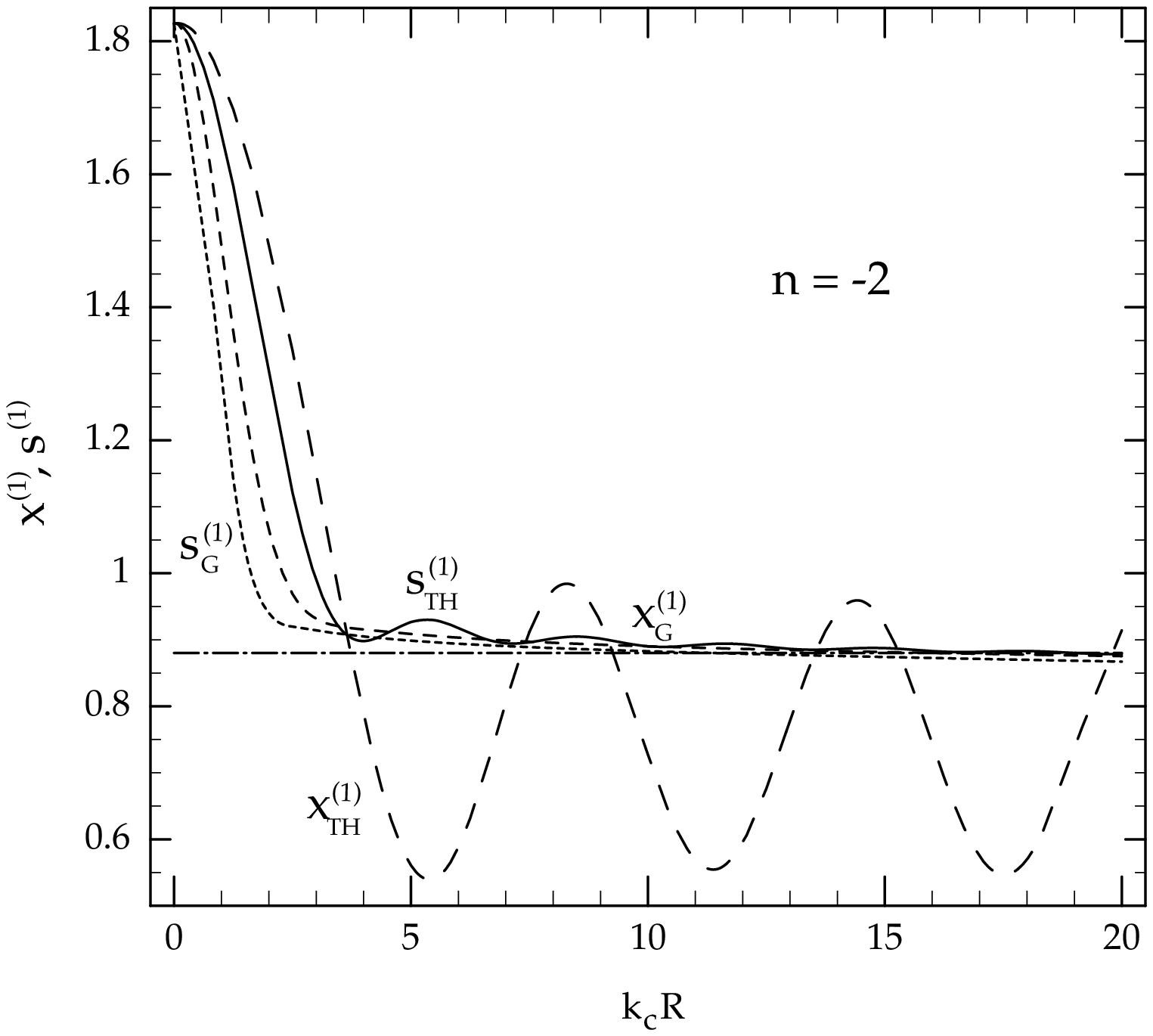}}

\caption{One-loop corrections to the average two-point correlation
function $x^{(1)}$ and variance $s^{(1)}$ for $n=-2$ as a function 
of $k_c R$, for top-hat (TH) and Gaussian (G) smoothing. Solid
curve shows $s_{\rm TH}^{(1)}$, dotted curve corresponds to
$s_{\rm G}^{(1)}$, dashed curve shows $x_{\rm G}^{(1)}$, and the
long-dashed curve $x_{\rm TH}^{(1)}$. The dot-dashed
line shows the large-scale approximation given in Eqs. ({\protect
\ref{s1infnm2th}}) and ({\protect \ref{s1infnm2g}}). 
For $k_c R=0$ we recover the unsmoothed results  $s^{(1)} =
x^{(1)} \approx 1.82$ ~({\protect \cite{ScFr96}}).}
\label{s1nm2}
\end{figure}

\begin{figure}[t!]
\centering
\centerline{\epsfxsize=10. truecm \epsfysize=9. truecm \epsfbox{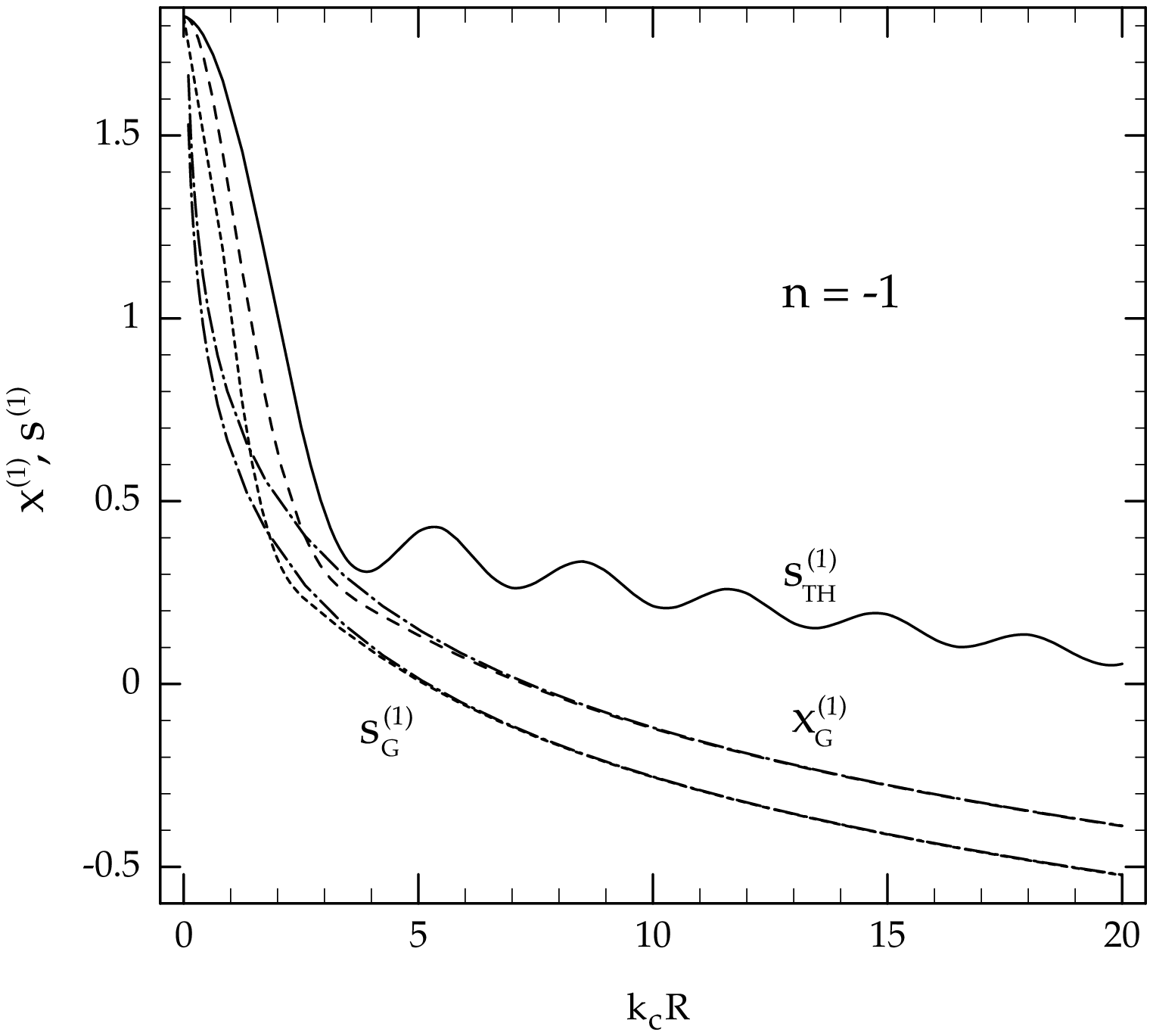}}

\caption{Same as  Fig.~{\protect \ref{s1nm2}} but for $n=-1$. 
Dot-dashed curves correspond to the large-scale approximation to the 
1-loop coefficients, Eqs.~({\protect \ref{s1infnm1g}}) 
and~({\protect \ref{x1infnm1g}}). We do not show 
$x^{(1)}_{\rm TH}$, which undergoes large oscillations.}
\label{s1nm1}
\end{figure}

\begin{figure}[t!]
\centering
\centerline{\epsfxsize=10. truecm \epsfysize=9. truecm \epsfbox{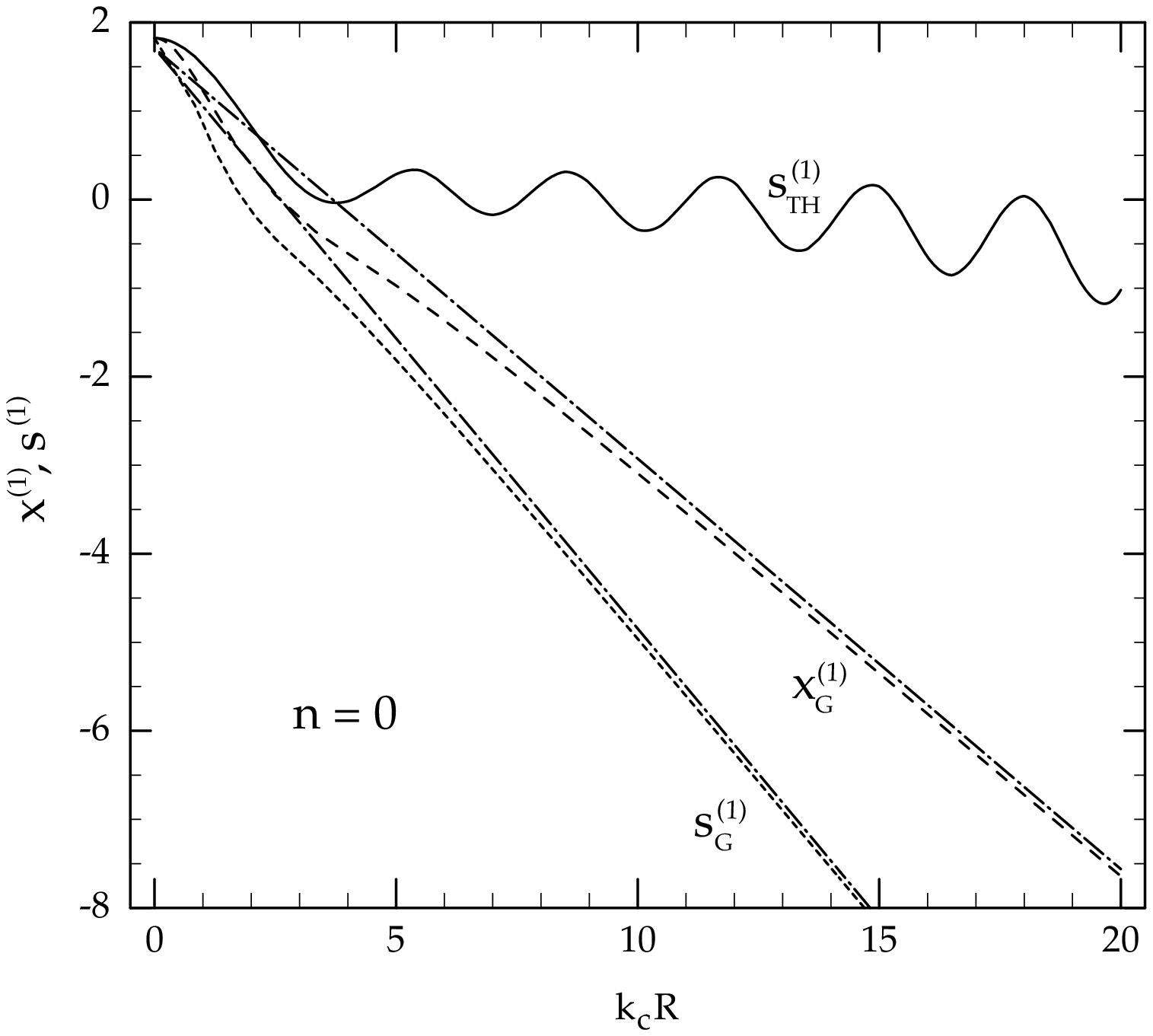}}

\caption{Same as  Fig.~{\protect \ref{s1nm2}} but for $n=0$. 
Dot-dashed curves correspond to the large-scale approximation to the  
1-loop coefficients, Eqs.~({\protect \ref{s1infn0g}}) 
and~({\protect \ref{x1infn0g}}).}
\label{s1n0}
\end{figure}

\begin{figure}[t!]
\centering
\centerline{\epsfxsize=10. truecm \epsfysize=9. truecm \epsfbox{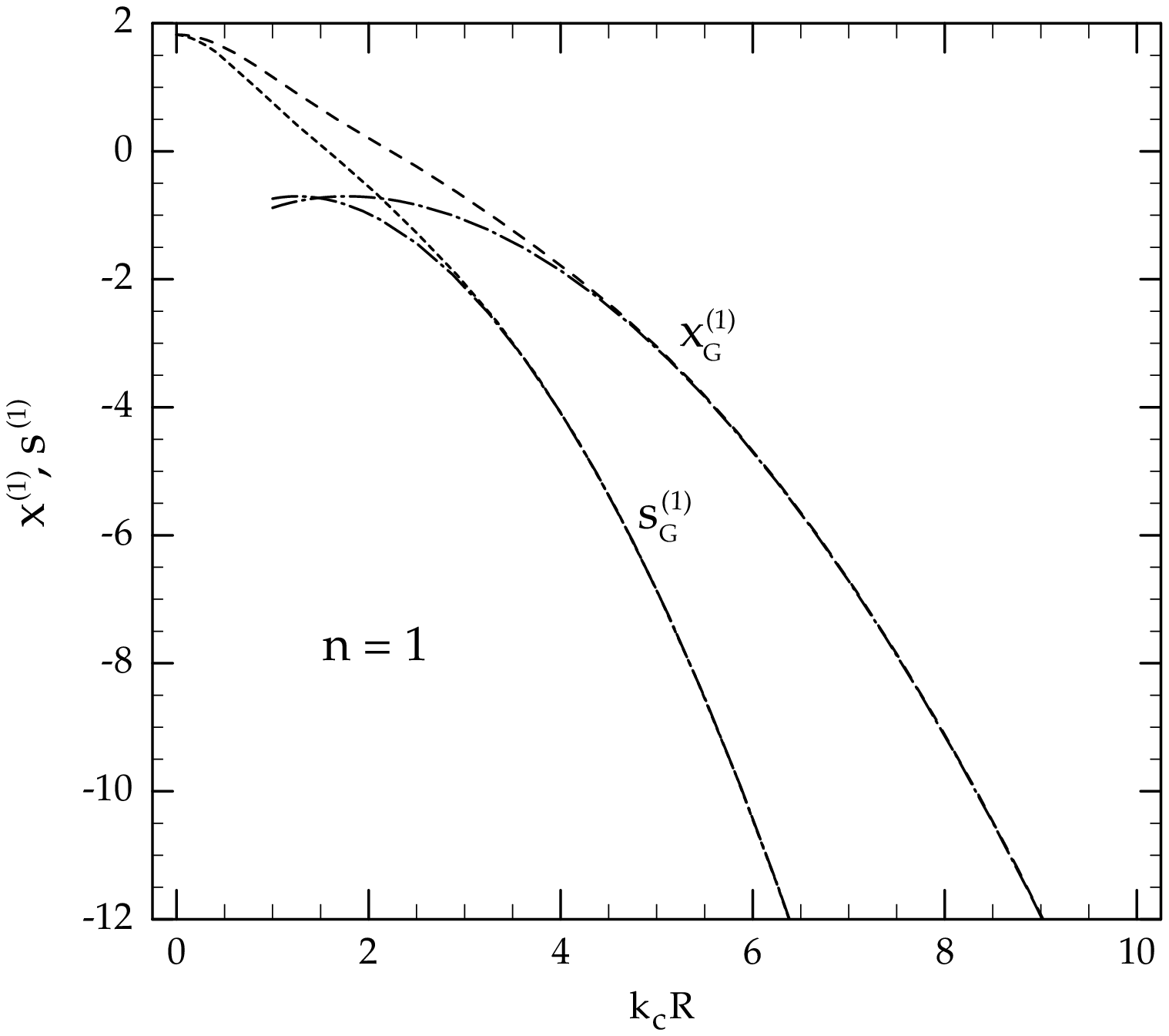}}

\caption{Same as  Fig.~{\protect \ref{s1nm2}} but for $n=1$. 
Dot-dashed curves correspond to the large-scale approximation to 
1-loop coefficients, Eqs.~({\protect \ref{s1infn1g}}) 
and~({\protect \ref{x1infn1g}}).}
\label{s1n1}
\end{figure}

In Figs.~\ref{s1nm2} - \ref{s1n1}, we 
show results for $s^{(1)}$ and $x^{(1)}$ for top-hat and Gaussian 
smoothing, based on numerical integration of Eq.~(\ref{s1x1ints}).
Similar results for $s^{(1)}$ for the Gaussian window function 
were presented by \L okas et al. (1995); comparison
with our results shows very good agreement
for scales such that $k_c R \ga 2$. However, in the limit 
$k_c R \rightarrow 0$, they apparently find  
$s^{(1)} \approx 0$, instead of recovering 
the unsmoothed values reported in Scoccimarro \& Frieman (1996)
 and in conflict with 
our results shown here. The source of their error may have been to 
replace $2k_c R$ by $k_c R$ in the upper limit of integration in 
Eq.~(\ref{s1int}); we found we could approximately reproduce  
their results with this change. 

From Eqs.~(\ref{s1x1ints}),  
one can analytically calculate $s^{(1)}$ and 
$x^{(1)}$ in the limit $k_c R \gg 1$ 
by asymptotic expansion in $1/k_c R$: for large $k_c R$,  all 
we need is the behavior of $p^{(1)}(n;x,\Lambda)$ for small $x$ given 
in the previous section.  For $n=-2$ we have:

\begin{equation}
	s^{(1)}(-2;k_cR\gg 1) \approx s^{(1)}(-2;\infty) = \frac{55 \pi^2}{392} 
	\frac{I_\sigma(-1,\infty)}{[I_\sigma(-2,\infty)]^2}
	\label{s1inftynm2},
\end{equation}

\noindent where we used Eqs.~(\ref{pm2x0}) and~(\ref{Isigma1st}). An 
identical expression holds for $x^{(1)}$ upon replacing $I_\sigma$'s by 
$I_\xi$'s. Using 
the results in Appendix~\ref{app:IsigIxi}, we find for {\it top-hat 
smoothing}:

\begin{eqnarray}
	s^{(1)}_{\rm TH}(-2;\infty) & = & \frac{1375}{1568} \approx 0.877
	\label{s1infnm2th},  \\
	x^{(1)}_{\rm TH}(-2;\infty) & = & \frac{110}{147} \approx 0.748
	\label{x1infnm2th}.
\end{eqnarray}

\noindent For {\it Gaussian smoothing} we find

\begin{equation}
	s^{(1)}_{\rm G}(-2;\infty) = x^{(1)}_{\rm G}(-2;\infty)
  =  \frac{55 \pi}{196} \approx 0.882
	\label{s1infnm2g}
\end{equation}

\noindent in good agreement with the numerical results of
\L okas et al. (1995), 
who found $s^{(1)}_{\rm G}(-2;\infty) \simeq 0.86$. 
Comparing these analytic results to Fig.~\ref{s1nm2}, we 
see excellent agreement except for $x^{(1)}_{\rm TH}$, 
which oscillates around the value given by Eq.~(\ref{x1infnm2th}). The 
reason for these oscillations can be partially understood from  
Eq.~(\ref{s1inftynm2}), using the results in
Appendix~\ref{app:IsigIxi} for finite $k_c R$. The behavior of
$x^{(1)}_{\rm TH}$ is also due in part to the fact that smoothing with 
one top-hat window allows small-scale power in 
$p^{(1)}(n;x,\Lambda)$ to `leak in' to the average correlation 
function, while ~(\ref{s1inftynm2}) assumes that only the large-scale 
power contributes. For this reason, in the following we 
will give analytical results only for Gaussian smoothing.

For $n=-1$, using Eqs.~(\ref{pm1x0}) and~(\ref{Isigma1st}) we have:

\begin{eqnarray}
	s^{(1)}(-1;k_cR\gg 1) & = &\frac{3068}{11025} 
	\frac{I_\sigma(1,\infty)}{[I_\sigma(-1,\infty)]^2} + \frac{244}{315}
	\int_0^\infty W^2(u)  u^3 \ln u \ du - \frac{61}{315}  
	\frac{I_\sigma(1,\infty)}{ [I_\sigma(-1,\infty)]^2} \nonumber \\
	& & \times \ln (k_c R)
	\label{s1inftynm1}
\end{eqnarray}

\noindent and similarly for $x^{(1)}$. Using Gaussian smoothing we find:


\begin{eqnarray}
	s^{(1)}_{\rm G}(-1;k_cR\gg 1) & = & 
\frac{6136}{11025}  + \frac{61}{315} (1-
	\gamma_e) - \frac{122}{315} \ln (k_c R) 
\approx  0.638 - 0.387 \ln (k_c R)
    \label{s1infnm1g},  \\
	x^{(1)}_{\rm G}(-1;k_cR\gg 1) & = &  \frac{6136}{11025}  + 
\frac{61}{315} (1-
	\gamma_e + \ln 2) - \frac{122}{315} \ln (k_c R) \approx 0.773 - 0.387
	\ln (k_c R) \nonumber \\
	\label{x1infnm1g}
\end{eqnarray}

\noindent where $\gamma_e \simeq  0.577216...$ is  the 
Euler-Mascheroni constant. 
Eqs.~(\ref{s1infnm1g}) and~(\ref{x1infnm1g}) are plotted as the dot-dashed 
curves in Fig.~\ref{s1nm1}, which shows the excellent agreement with the 
results from the full numerical integration at large $k_c R$.

For $n=0$, one must include the next to leading order term at small $x$ in 
$p^{(1)}(0;x,\Lambda)$, since this gives rise to a constant term in the 
1-loop coefficients. We have:

\begin{equation}
    p^{(1)}(0;x) \approx - \frac{244}{315} x^2 + \frac{20 \pi^2}{147} x^3 
    + {\cal O} (x^4)
	\label{p1n0ntl},
\end{equation}

\noindent which gives (see dot-dashed curves in Fig.~\ref{s1n0}):

\begin{eqnarray}
	s^{(1)}_{\rm G}(0;k_cR\gg 1) & = & \frac{80 \pi}{147}  - \frac{122}{105 \sqrt{\pi}} 
	\  k_c R \approx 1.710 - 0.655 \ k_c R
    \label{s1infn0g},  \\
	x^{(1)}_{\rm G}(0;k_cR\gg 1) & = &  \frac{80 \pi}{147}  - \frac{61}{105}
	\sqrt{\frac{2}{\pi}} 
	\  k_c R \approx 1.710 - 0.464 \ k_c R
	\label{x1infn0g}.
\end{eqnarray}

Similarly, for $n=1$, we need the expansion:

\begin{equation}
    p^{(1)}(1;x) \approx - \frac{122}{315} x^3 + \frac{18}{49} x^4 
    - \Big( \frac{4973}{22050} + \frac{16}{105} \ln x \Big)\ x^5 
    + {\cal O} (x^6)
	\label{p1n1ntl},
\end{equation}

\noindent which leads to (see Fig.~\ref{s1n1}):

\begin{eqnarray}
	s^{(1)}_{\rm G}(1;k_cR\gg 1) & = & -\frac{4973}{7350}  - 
\frac{4}{105} (11 - 6 \gamma_e)
	+ \frac{16}{35} \ln (k_c R) + \frac{135 \sqrt{\pi}}{392} k_c R -
	\frac{122}{315} (k_c R)^2 \nonumber \\
& \approx & -0.964 + 0.457 \ln (k_c R) + 0.610 \ k_c R - 0.387 \ (k_c R)^2
    \label{s1infn1g},  
\end{eqnarray}
\begin{eqnarray}
	x^{(1)}_{\rm G}(1;k_cR\gg 1) & = &  -\frac{4973}{7350}  - 
\frac{4}{105} (11 - 6 \gamma_e + 6 \ln 2)
	+ \frac{16}{35} \ln (k_c R) + \frac{135 }{392} \sqrt{\frac{\pi}{2}} 
k_c R  \nonumber \\
& & -	\frac{61}{315} (k_c R)^2 
\approx -1.122 + 0.457 \ln (k_c R) + 0.432 \ k_c R - 0.194 \ (k_c R)^2. \nonumber \\
& & 	\label{x1infn1g}
\end{eqnarray}

Note that the 1-loop coefficients are scale-dependent 
for $n \geq -1$; this is an indication of the breaking of  
self-similarity, as we discuss next. 
(A similar analysis to that above 
can be carried out for the 1-loop corrections in the 
Zel'dovich approximation, but details will not be given here.)

\section{Self-Similarity and Perturbation Theory} 
\label{sec:selfsimpt}

\subsection{Self-Similar Solutions}
\label{sec:selfsimsol}

Since there is 
no preferred scale in the dynamics of a self-gravitating  
pressureless perfect fluid in an 
Einstein-de Sitter universe,  
Eqns.~(\ref{eqsmotionrealspace}) admit self-similar solutions (see 
\cite{Peebles80}). 
This means that the cosmological fields 
should scale with a self-similarity variable, given appropriate initial 
conditions: knowing the fields  
at a given time completely specifies their evolution. We can search 
for the appropriate self-similarity transformation by rewriting the 
fields as

\begin{mathletters}
	\label{ssansatz}
 \begin{eqnarray}
 	\delta ({\bf x}, \tau) & \equiv & \chi ({\bf y})
 	\label{deltass},  \\
 	{\bf v} ({\bf x}, \tau) & \equiv & \tau^\mu {\bf u} ({\bf y})
 	\label{vss},  \\
 	\Phi ({\bf x}, \tau) & \equiv & \tau^\gamma \varphi ({\bf y})
 	\label{phiss},
 \end{eqnarray}
 \end{mathletters}
 
\noindent where ${\bf y} \equiv {\bf x}/ \tau^\nu$ is the similarity variable. 
Self-similarity is obeyed if we can rewrite Eqs.~(\ref{eqsmotionrealspace}) in 
terms  of ${\bf y}$ only; this condition determines the indices $\mu$, 
$\gamma$, and $\nu$.  
Note that the term $1+\delta{({\bf x},\tau)}$ 
in the continuity equation~(\ref{continuity}) precludes a power of $\tau$ 
prefactor in Eq.~(\ref{deltass}). Substituting the ansatz  
Eq.~(\ref{ssansatz}) into Eq.~(\ref{continuity}) and equating powers of 
$\tau$ we get the condition

\begin{equation}
	\mu= \nu -1
	\label{betass},
\end{equation}

\noindent whereas the Poisson equation Eq.~(\ref{poisson}) gives:

\begin{equation}
	\gamma = 2 (\nu -1)
	\label{gammass}.
\end{equation}

\noindent These two conditions in turn guarantee that the Euler equation 
obeys self-similar evolution. The index $\nu$ 
can be determined in terms of the spectral index $n$ of the linear power  
spectrum: Eq.~(\ref{deltass}) implies that the two-point 
correlation function obeys the self-similar scaling

\begin{equation}
		\xi ( x, \tau)  \equiv \Xi ( x \tau^{-\nu})
	\label{xiss},
\end{equation}
  
  \noindent which upon Fourier transformation gives for the power 
spectrum   
  
\begin{equation}
		P (k, \tau) \equiv  \tau^{3 \nu} \ {\cal P} ( k \tau^\nu)
	\label{pss}.
\end{equation}
  
\noindent On the other hand, in the linear regime,  
for a power-law initial power spectrum with spectral index $n$ 
in an Einstein-de Sitter universe, $P_{11}(k,\tau) \sim \tau^4 k^n$; 
equating this with (\ref{pss}) fixes the remaining parameter in the 
similarity transformation,

\begin{equation}
	\nu = \frac{4}{n+3}
	\label{alphass},
\end{equation}

\noindent (A more general derivation of (\ref{alphass}) starts from   
the BBGKY hierarchy of equations for 
the phase space particle distribution functions,~\cite{DP77}.)  
Using these results, the equations of motion   
(\ref{eqsmotionrealspace}) can be rewritten in self-similar form:

\begin{mathletters}
\label{eqsmotionrealspacess}

\begin{equation}
	\nu   {\bf y} \cdot \hat \nabla \chi({\bf y}) - \hat \nabla \cdot  \{
	[1+\chi {({\bf y})}] {\bf u}({\bf y}) \} =  0
	\label{continuityss},
\end{equation}
\begin{equation}
     \nu  {\bf y} \cdot \hat \nabla {\bf u}({\bf y}) - 
     (1+\nu)  {\bf u}({\bf y})
     - [{\bf u}({\bf y}) \cdot \hat \nabla] {\bf u}({\bf y}) =   \hat \nabla 
	\varphi({\bf y})
	\label{eulerss},
\end{equation}	
\begin{equation}
	\hat \nabla^2 \varphi({\bf y}) = 6 \chi({\bf y})
	\label{poissonss},
\end{equation}
 \end{mathletters}	 

\noindent where $\hat \nabla \equiv \partial / \partial {\bf y}$. 
The solutions of these equations in the linear regime
are $\chi({\bf y}) \propto y^\zeta$ with $\zeta= -2/\nu, 3/\nu$, which  
correspond to the well-known linear growing and decaying modes,  
$\delta_1 \propto a,a^{-3/2}$. The condition of 
self-similar evolution also fixes the spatial dependence of 
the fields, e.g., $\delta_1({\bf x}) \propto x^{-2/\nu}$ for the growing 
mode. However, we generally consider the density and velocity fields to 
be random fields, on which we impose initial conditions only 
{\it statistically}, e.g., by specifying 
the linear power spectrum for Gaussian initial conditions. A particular 
realization of the ensemble of initial conditions will not obey  
this spatial scaling, so the cosmological fields themselves 
will not be self-similar (\cite{JaBe95}).  
Since NLCPT is built from linear 
solutions, the same conclusion applies to the 
expansions given by
Eq.~(\ref{ptansatz}), which 
are not self-similar (except when the spatial dependence is fixed as above).
Nevertheless, in some cases the  
{\it statistical} quantities of interest, such as the power spectrum, 
variance, and two-point correlation function, may exhibit self-similar 
scaling even if the fields in a given realization do not. We now 
consider the conditions under which this happens in perturbation theory.

\subsection{Self-Similarity and Linear Perturbation Theory}
\label{sec:selfsimLPT}

The introduction of fixed (time-independent) 
cutoff scales $\epsilon$ and $k_c$ 
in the linear power spectrum (\ref{P11}) breaks 
self-similarity, because they do not scale with the 
self-similarity variable $k a^{2/(n+3)}$ (the Fourier space analog 
of $y$). The 
extent to which one can take the limits $\epsilon 
\rightarrow 0$ and $k_c \rightarrow \infty$ will determine whether 
the statistical properties of the density field obey  
self-similar scaling.  
In the absence of cutoffs, the only physical 
scale that can be defined from the power spectrum is 
the correlation length, $R_0$, defined by:

\begin{equation}
	\sigma^2(R_0) = \int P(k,\tau) W^2(k R_0) d^3 k \equiv 1
	\label{corrlength1}.
\end{equation}

\noindent In linear perturbation theory, for scale-free initial conditions, 
one finds $R_0 \propto a^{2/(n+3)}$, 
which has the right time dependence to build the
self-similarity variable. Consequently, in the {\it absence} of cutoffs,
statistical quantities in linear theory evolve self-similarly
with $R_0$, and we can write, e.g., 
(see Eqs.~(\ref{xiss})~and~(\ref{pss})):

\begin{equation}
		\xi_\ell(r, \tau)  \equiv \Xi_\ell (r/R_0)
	\label{xiss2},
\end{equation}
    
\begin{equation}
	R_0^{-3} P_{11}(k, \tau) \equiv  {\cal P}_{11} ( k R_0)
	\label{pss2}.
\end{equation}

The presence of  cutoffs in the initial spectrum changes this situation. 
 From Eq.~(\ref{P11}) we have

\begin{eqnarray}
	R_0^{n+3} & = & 4 \pi A a^2 \int_{\epsilon R_0}^{k_c R_0} \kappa^{n+2} 
	W^2(\kappa) d \kappa \equiv 4 \pi A a^2 [I_\sigma (n,k_c R_0) -
	I_\sigma (n, \epsilon R_0) ] \nonumber \\ & 
\equiv & 4 \pi A a^2 
	\Delta I_\sigma (n, k_c R_0,\Lambda)
	\label{r0},
\end{eqnarray}

\noindent and $R_0$ will not scale as  $a^{2/(n+3)}$ unless
$\Delta I_\sigma (n, k_c R_0,\Lambda)$ is time independent. Before
examining under which conditions this is true, it is useful to 
introduce another measure $\ell_0$ of the 
correlation length, defined in terms of the volume-averaged 
correlation function, 

\begin{equation}
	\bar \xi(\ell_0) = \int P(k,\tau) W(k \ell_0) d^3 k \equiv 1
	\label{corrlength2},
\end{equation}

\noindent and therefore: 

\begin{equation}
	\ell_0^{n+3} = 4 \pi A a^2 \int_{\epsilon \ell_0}^{k_c \ell_0} \kappa^{n+2} 
	W(\kappa) d \kappa \equiv 4 \pi A a^2 [I_\xi  (n,k_c \ell_0) -
	I_\xi (n, \epsilon  \ell_0) ] \equiv 4 \pi A a^2 
	\Delta I_\xi (n, k_c \ell_0,\Lambda)
	\label{l0}.
\end{equation}

\noindent Note that $R_0$ and $\ell_0$ both differ 
slightly from the conventional definition of the correlation length,  
$r_0$, in terms of the two-point correlation function, $\xi(r_0)\equiv 1$.

\begin{figure}[t!]
\centering
\centerline{\epsfxsize=11. truecm \epsfysize=8. truecm \epsfbox{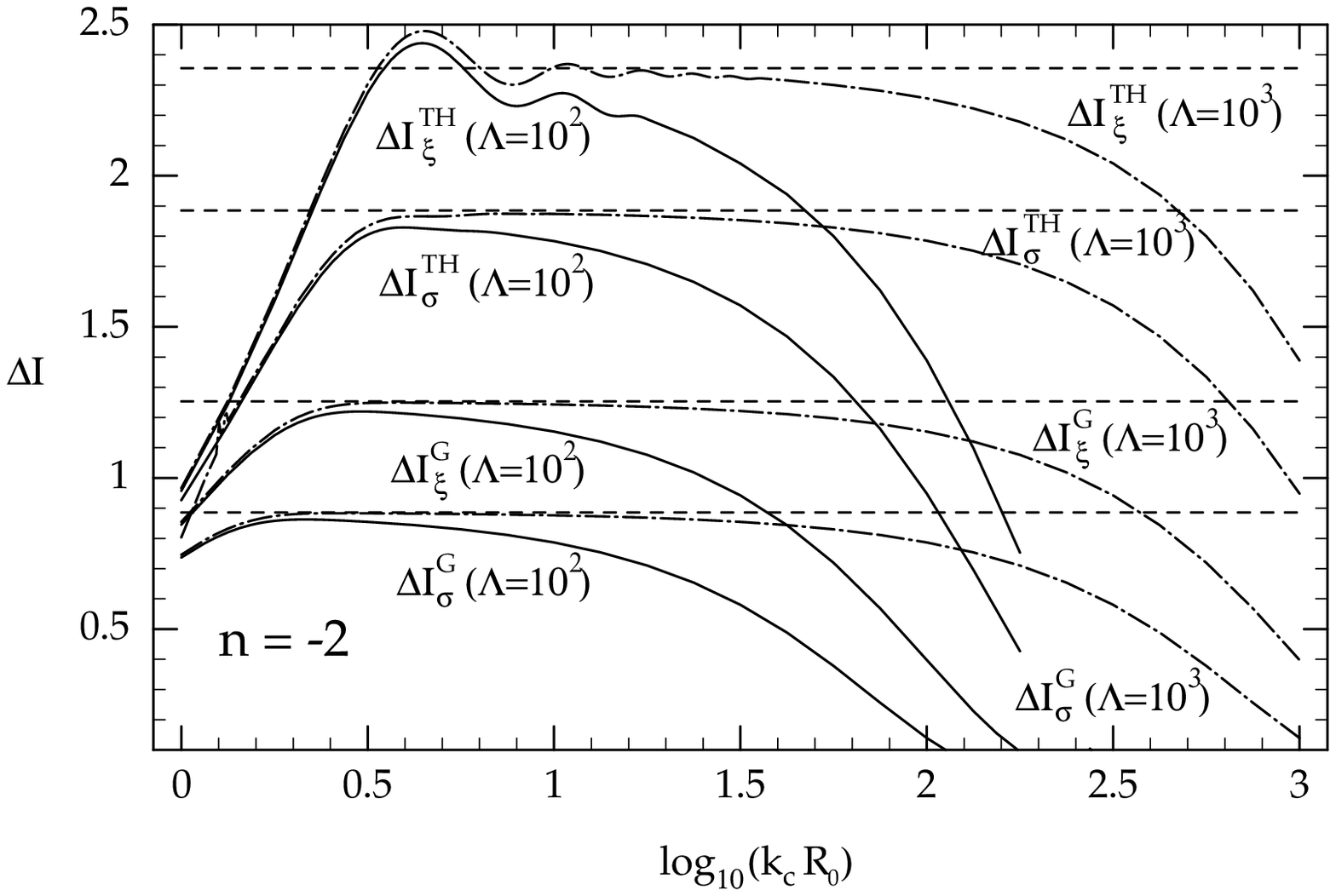}}
\caption{$\Delta I_{\sigma,\xi}(-2, k_c R_0,\Lambda)$ as a 
function of $k_c R_0$ for top-hat and Gaussian smoothing (see  
Eqs.~(\protect \ref{r0}) and~(\protect \ref{l0}) for definitions).
 Dashed lines represent 
the self-similar case (constant $\Delta I$) 
in which the limit $\Lambda \equiv k_c / \epsilon \rightarrow \infty$ has 
been taken ($\epsilon \rightarrow 0$, $ k_c  \rightarrow \infty$). 
Dot-dashed curves denote the choice $\Lambda = 10^3$, solid  
curves correspond to $\Lambda = 10^2$. The differences between 
these curves and 
the  $\Lambda \rightarrow \infty$ lines  
measure the deviations of the correlation
length from self-similar evolution.}
\label{DInm2}
\end{figure}

A necessary condition for self-similarity 
in the variable $R_0$ ($\ell_0$) 
is the convergence of the integrals $\Delta I_\sigma $ ($\Delta I_\xi $) 
for small and large $\kappa$ (where $\kappa$ denotes $kR_0$ ($k\ell_0$));  
otherwise these quantities would be sensitive to the cutoffs and 
consequently time dependent.  
From Eqs.~(\ref{r0}) and~(\ref{l0}), in order for the 
correlation length to scale  self-similarly in linear theory, 
we must have {\it at least} the following conditions:

 \begin{enumerate}
 	\item  For a Gaussian filter, since $W(\kappa) \approx 1$ when $\kappa
 	 \rightarrow 0$, convergence in the infrared (small 
 	$\kappa$) requires $n > -3$. Convergence in the ultraviolet (large 
 	$\kappa$) is achieved for any $n$.
 
 	\item  For a top-hat filter, convergence in the 
 	infrared requires $n > -3$. Since  
 	$W_{\rm TH}(\kappa) \approx \kappa^{-2}$ 
 	as $\kappa \rightarrow \infty $, convergence in the ultraviolet 
 	requires $n < 1$ in Eq.~(\ref{r0}) and $n < -1$ in Eq.~(\ref{l0}).
 \end{enumerate} 

\noindent Note the dependence of these conditions 
on the local averaging scheme (top-hat vs. Gaussian)
and the type of statistics (variance vs. average correlation function).

\begin{figure}[t!]
\centering
\centerline{\epsfxsize=11. truecm \epsfysize=8. truecm \epsfbox{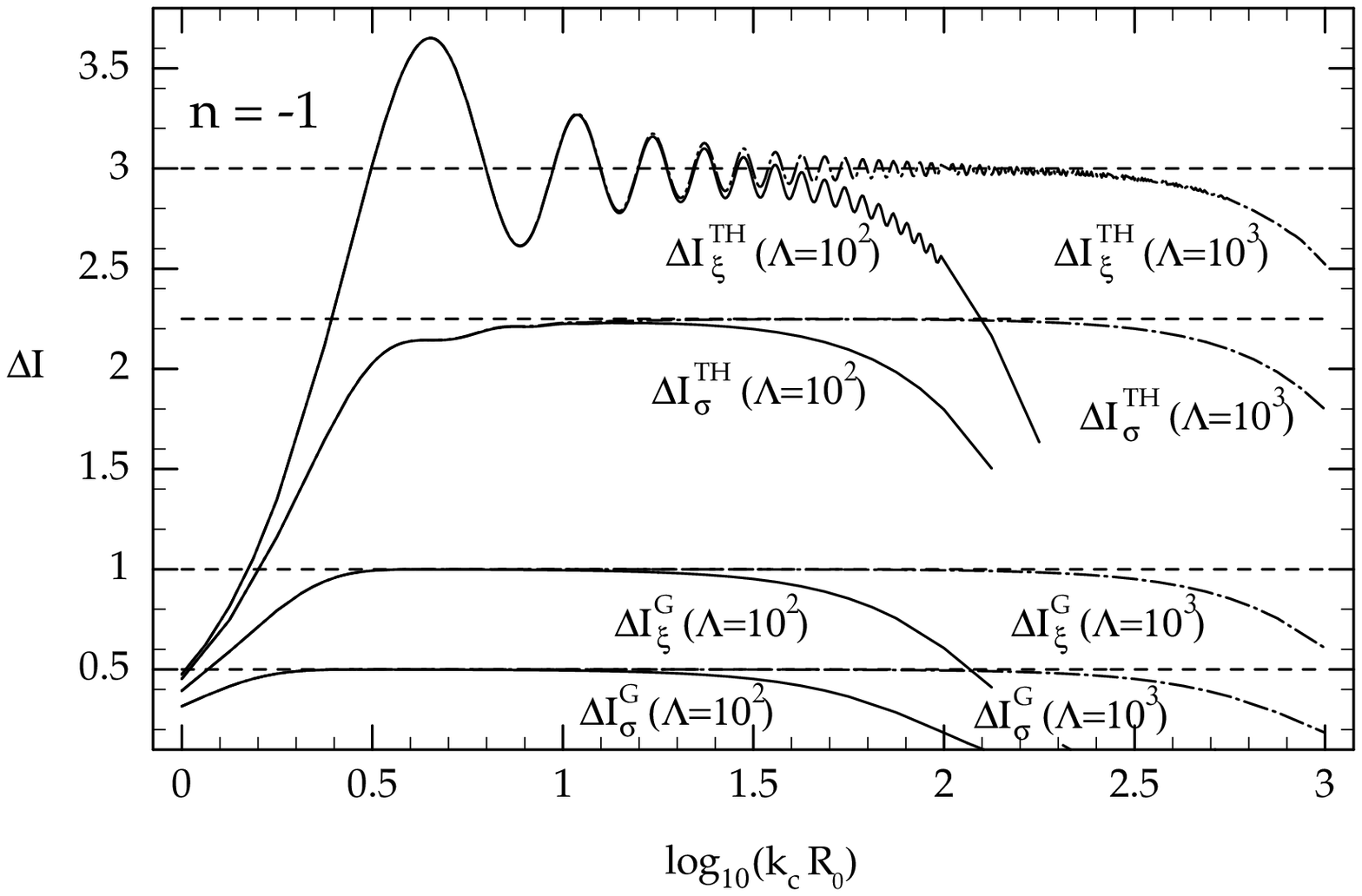}}

\caption{Same as Fig.~{\protect \ref{DInm2}} but for $n=-1$.}
\label{DInm1}
\end{figure}

\begin{figure}[t!]
\centering
\centerline{\epsfxsize=11. truecm \epsfysize=8. truecm \epsfbox{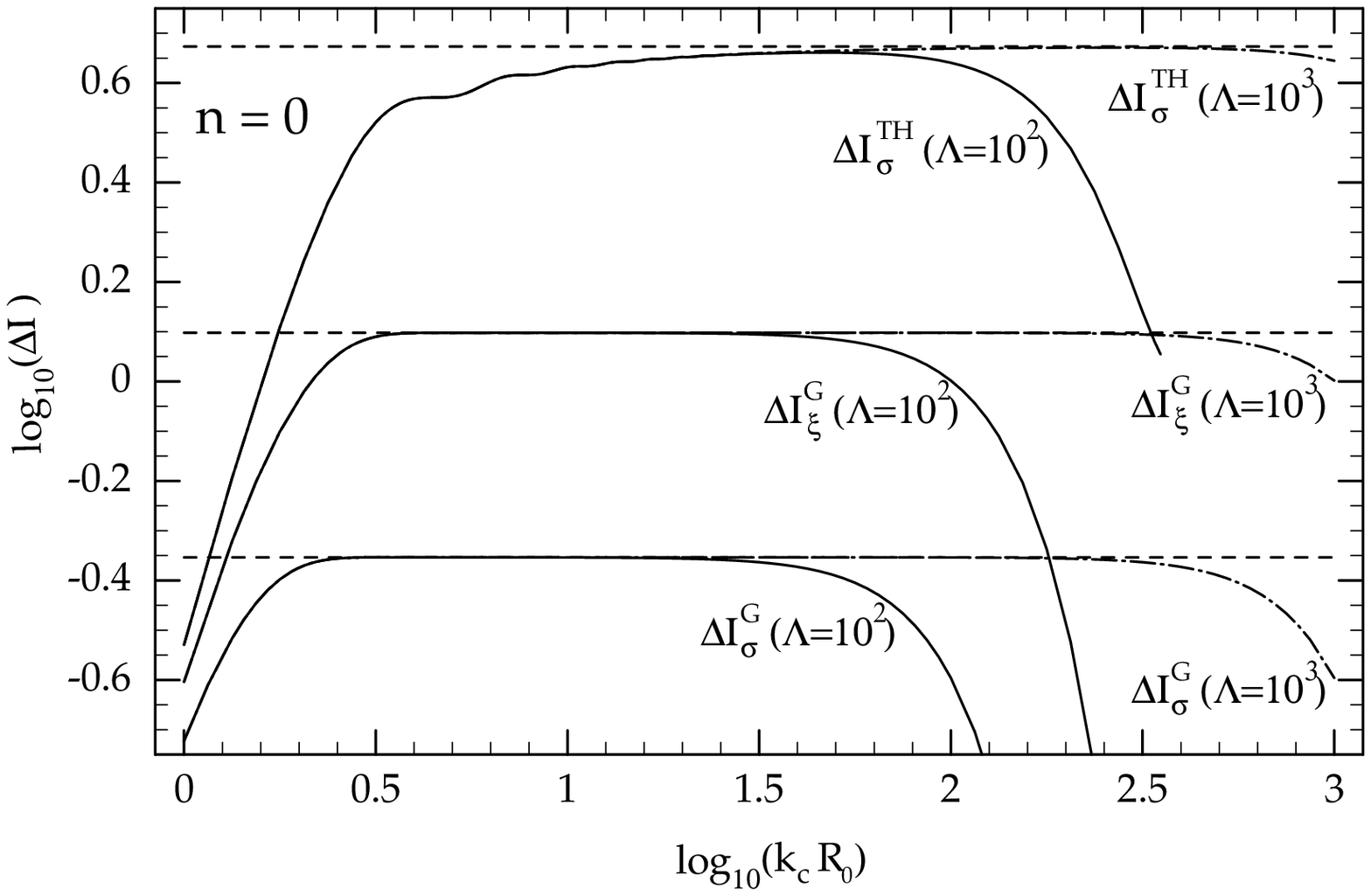}}

\caption{Same as Fig.~{\protect \ref{DInm2}} but for $n=0$. }
\label{DIn0}
\end{figure}

\begin{figure}[t!]
\centering
\centerline{\epsfxsize=11. truecm \epsfysize=8. truecm \epsfbox{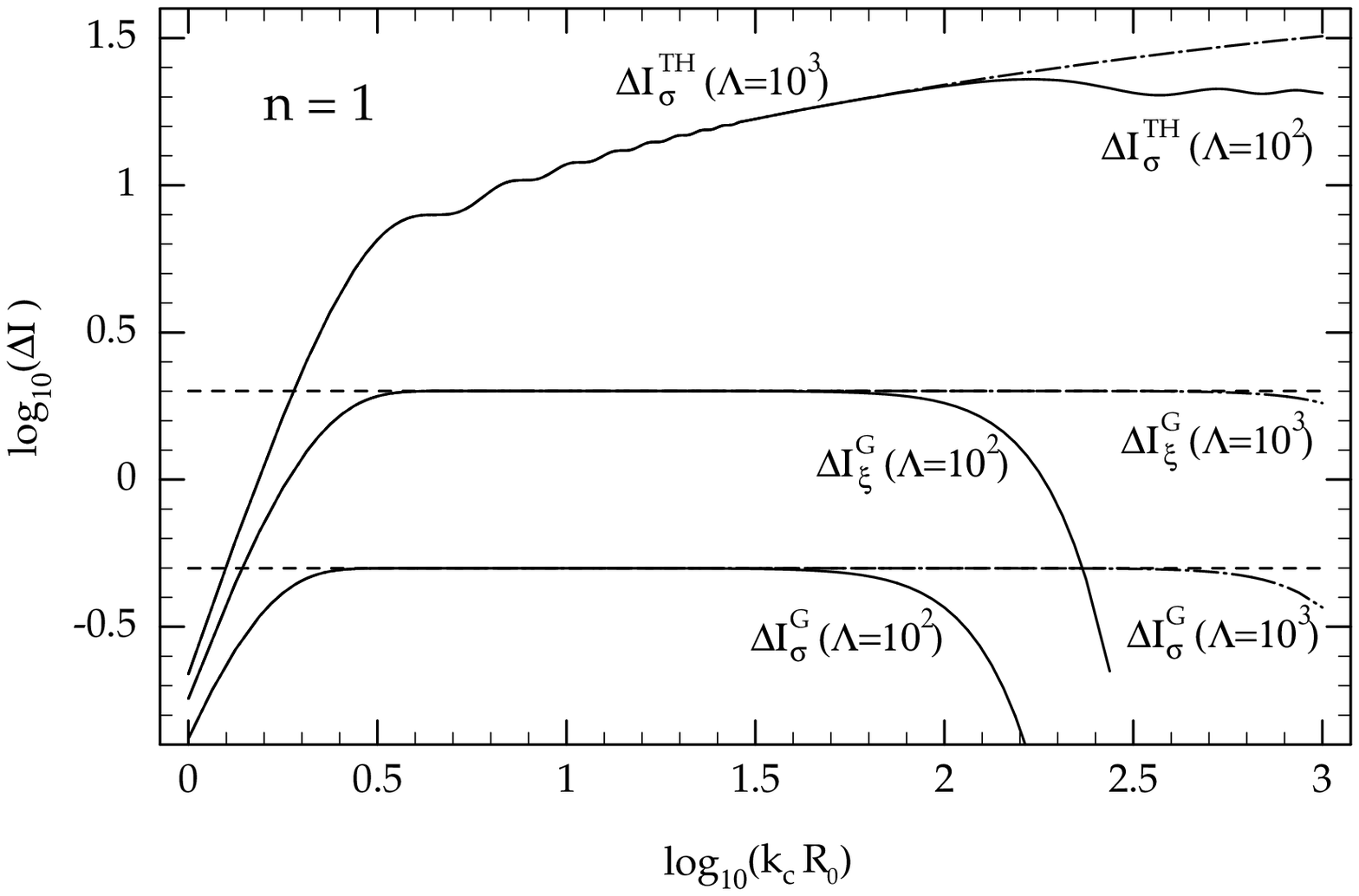}}

\caption{Same as Fig.~{\protect \ref{DInm2}} but for $n=1$.}
\label{DIn1}
\end{figure}

While these conditions guarantee convergence of $\Delta I_{\sigma,\xi}$, 
they do not test the constancy of these quantities for different
choices of parameters. In Figs.~\ref{DInm2} - \ref{DIn1}, 
we show $\Delta I_{\sigma,\xi}(n, k_c R_0,\Lambda)$ for 
$\Lambda \equiv k_c/\epsilon = 
10^2, 10^3, \infty$, as a function of $k_cR_0$ for  
spectral indices $n=-2,-1,0,1$. These finite values of $\Lambda$ are 
comparable to the dynamic range currently achievable in N-body simulations. 
By definition, $k_c R_0$ 
measures the ratio of the correlation length to the small scale cutoff of 
the linear power spectrum. Therefore, in the region $k_c R_0 \approx 1$, 
the evolution of the correlation length is strongly 
affected by the cutoff $k_c$, 
and the $\Delta I_{\sigma,\xi}$ drop precipitously. By the 
time the correlation length has evolved to $k_c R_0 \ga 3$, however, 
$\Delta I_{\sigma,\xi}$ reach their no-cutoff limits (in the cases where 
they converge). On the other hand, when $k_c R_0 \approx \Lambda$, i.e., 
$R_0 \approx \epsilon^{-1}$, the 
correlation length reaches the 
large-scale cutoff, and $\Delta 
I_{\sigma,\xi}$ again drops. This behavior is more dramatic as $n$ decreases 
(and therefore strongest for $n=-2$) due to the increase in large-scale 
power. As expected from the discussion above, 
the definition of the correlation length using 
Gaussian smoothing displays self-similar evolution 
over a longer time interval than the 
other possibilities, with top-hat smoothing becoming notoriously worse as 
$n$ increases. 

The behavior of statistical quantities in linear perturbation theory 
with respect to self-similarity is determined directly from these 
considerations. We can write the linear power spectrum as (see 
Eq.~(\ref{pss2}) ):

\begin{equation}
	{\cal P}_{11}(k R_0,k_c R_0,\Lambda) \equiv \frac{(k R_0)^n}{4 \pi 
	\Delta I(n,k_c R_0,\Lambda)} 
	\label{ssp11lin},
\end{equation}

\noindent where $\Delta I$ depends on the local averaging scheme and 
statistical quantity used to define the correlation length, and 
$k R_0 \ll  1$ in the linear regime. Unless 
otherwise noted, for power spectrum calculations 
we will use the choice in Eq.~(\ref{r0}), i.e., $R_0$,  
with Gaussian smoothing. For the  variance 
and average correlation function we have

\begin{equation}
	\sigma^2_\ell(R/R_0,k_c R_0,\Lambda) \equiv 
	\frac{\Delta I_\sigma(n,k_c R,\Lambda)}
	{\Delta I_\sigma(n,k_c R_0,\Lambda)} \ 
	\left(\frac{R}{R_0}\right)^{-(n+3)}
	\label{sssiglin},
\end{equation}

\begin{equation}
	\bar \xi_\ell(R/R_0,k_c R_0,\Lambda) \equiv 
	\frac{\Delta I_\xi(n,k_c R,\Lambda)}
	{\Delta I_\xi(n,k_c R_0,\Lambda)} \ 
	\left(\frac{R}{R_0}\right)^{-(n+3)}
	\label{ssxilin}.
\end{equation}

\noindent Now the question is under what conditions are the statistics  
self-similar in the variables $k R_0$ (for the power spectrum) 
and $R/R_0$ (variance and 
average correlation function). In the regime $R \gg R_0$ 
(where linear theory is valid) and 

\begin{equation}
	3 \la k_c R_0 \ll k_c R \ll \Lambda 
	\label{sscond},
\end{equation}

\noindent Figs.~\ref{DInm2} - \ref{DIn1} show that 
$\Delta I $ is approximately constant. Provided the conditions 
on $n$ stated above hold, in this regime 
$k R_0$ and $R/R_0$ are self-similar variables: 
in linear perturbation theory, the power spectrum scales as 

\begin{mathletters}
\begin{equation}
	{\cal P}_{11}(k R_0,k_c R_0,\Lambda) \approx  \frac{(k R_0)^n}{2 \pi 
	\Gamma \left(\frac{n+3}{2}\right)} 
	\label{p11ssregion},
\end{equation}

\noindent (see Eq.~(\ref{IsiginftyG})), and

\begin{equation}
	\sigma^2_\ell(R/R_0,k_c R_0,\Lambda) \approx 
	\bar \xi_\ell(R/R_0,k_c R_0,\Lambda) \approx
	 \left(\frac{R}{R_0}\right)^{-(n+3)}
	\label{sigxissregion}.
\end{equation}
\end{mathletters}

\noindent Note that, for top-hat smoothing, the requirement of 
self-similarity of the statistical quantities is in practice
less restrictive than the requirement that $R_0$ or $\ell_0$ scale 
self-similarly.
For example, although $\ell_0$, defined from top-hat smoothing, requires 
$n \la -1$ to scale 
self-similarly, the average correlation function for $n=1$ {\it does}  
scale self-similarly, i.e., it depends on $R/R_0$ only, 
in the regime given by
Eq.~(\ref{sscond}),  

\begin{equation}
	\bar \xi_\ell(R/R_0,k_c R_0,\Lambda) \approx  
\left(\frac{R}{R_0}\right)^{-3}
	\label{ssxilinn1},
\end{equation}

\noindent (see Eq.~(\ref{Ixin1TH})) 
where we have averaged over oscillatory behavior. 
However, it does not obey the ``expected" $\bar 
\xi_\ell(R/R_0,k_c R_0,\Lambda) 
\approx  (R/R_0)^{-4}$ scaling. This trend appears to be seen in the results
from high-resolution scale-free N-body simulations~(\cite{CBH95}).


\subsection{Self-Similarity and Non-Linear Perturbation Theory}
\label{sec:selfsimNLCPT}

We have examined the conditions under which linear perturbation 
theory  obeys self-similar scaling. 
We now turn to the question of whether loop corrections in NLCPT
affect self-similar scaling.  

Using the results of the previous section and Eq.~(\ref{P2def}), we can 
write the power spectrum up to 1-loop corrections as:

\begin{equation}
	{\cal P}(k R_0,k_c R_0,\Lambda) \equiv \frac{(k R_0)^n}{4 \pi 
	\Delta I_\sigma(n,k_c R_0,\Lambda)}  + \frac{(k_c R_0)^{2n+3}}{16 \pi 
	[\Delta I_\sigma(n,k_c R_0,\Lambda)]^2} \ p^{(1)}(n;k / k_c,
	\Lambda)
	\label{ssp1l}, 
\end{equation}

\noindent where $kR_0 \la 1$ in the weakly non-linear regime.   
Self-similarity 
is maintained at the 1-loop level if, in the scaling regime given by  

\begin{equation}
	k R_0 \la  1 \ll k_c R_0 \ll \Lambda
	\label{ssreg},
\end{equation}

\noindent the dependence of ${\cal P}(k R_0,k_c R_0,\Lambda)$ 
on $k_c R_0$ and $\Lambda$ is negligible, that is, if

\begin{equation}
	p^{(1)}(n;k / k_c \rightarrow 0,\Lambda \rightarrow \infty)
	\approx a_n \left( \frac{k}{k_c } \right)^{2n+3}
	\label{sscondp1l},
\end{equation}

\noindent with $a_n$ some constant which depends only on the spectral 
index $n$. From the results of Section~\ref{sec:1lps}, we see that
this condition is only satisfied when $n=-2$. In this case we have (see 
Eq.~(\ref{pm2x0})):

\begin{equation}
	{\cal P}(k R_0;n=-2) \approx \frac{1}{2 \pi^{3/2}} (k R_0)^{-2}  + 
	\frac{55  }{392} (k R_0)^{-1}
	\label{ssp1lnm2},
\end{equation}

\noindent which takes the self-similar form. For $n=-1$, 1-loop 
diagrams yield logarithmic $k_c$-dependent corrections 
to self-similar scaling (see Eq.~(\ref{pm1x0})),

\begin{equation}
	{\cal P}(k R_0,k_c R_0) \approx \frac{1}{2 \pi} (k R_0)^{-1}  + 
	\frac{16}{11025 \pi} \ (k R_0) \ \left[ 3068 + 2135 \ln \left( 
	\frac{k}{k_c} \right) \right]
	\label{ssp1lnm1}.
\end{equation}

\noindent The results for $n = 0,1$ show a stronger power law breaking of 
self-similarity (see 
Eqs.~(\ref{f0x0}) and (\ref{f1x0})):

\begin{equation}
	{\cal P}(k R_0,k_c R_0) \approx \frac{(k R_0)^{n}}{2 \pi 
	\ \Gamma \left( \frac{n+3}{2} \right)}  - \frac{61 \ (k R_0)^{2n+3}}
	{315 \pi  (n+1) \  \Gamma^2 \left( \frac{n+3}{2} \right) } \ \left( 
	\frac{k}{k_c} \right)^{\eta}
	\label{ssp1lngeq0},
\end{equation}

\noindent where $\eta = -(n+1)$ is an exponent which measures the deviation 
from self-similar scaling, generally known as {\it the anomalous 
dimension} in the theory of critical 
phenomena~(\cite{Goldenfeld92,Barenblatt79}). 
A visual summary of these 
results is given in the next subsection, where we compare 1-loop NLCPT to the 
universal scaling hypothesis first proposed by Hamilton et
al. (1991). For the variance and average two-point
correlation function, self-similarity breaking means that, in the expected
scaling region given by Eq.~(\ref{sscond}), the 1-loop coefficients 
$s^{(1)}$ and $x^{(1)}$ (see Eq.~(\ref{sigxiexp}) ) are scale-dependent;  
in Sec.~\ref{sec:1lsv}, we found this to be the case
when $n \geq -1$. Thus, for $n \geq -1$, 
we find generally that self-similarity is 
broken in perturbation theory by the first non-linear (1-loop) corrections to 
the power spectrum.  

Our conclusions about self-similarity in NLCPT differ substantially 
from those of Jain \& Bertschinger (1995). The primary  
difference stems from the fact that they take 
the small scale cutoff 
in the linear power spectrum, $k_c$, to be {\it time dependent} in comoving 
coordinates. They choose $k_c$  
to scale as the inverse correlation length, i.e., they   
fix $k_c R_0 \approx 1$ throughout the 
evolution. Their rationale for this is plausible: since   
NLCPT is expected to break down at scales below the 
correlation length $R_0$, within the framework of perturbation theory 
one should perhaps restrict contributions to the  power spectrum 
to scales larger than $R_0$. Since $R_0$  
depends on time, the cutoff of the linear power spectrum 
must be chosen to scale as $1/R_0$ 
in order to satisfy this requirement at all times.
Using this condition on $k_c R_0$, they correctly conclude that the power 
spectrum evolves self-similarly (to any order in NLCPT), {\it provided}  
there are no infrared divergences. At 1-loop, we 
can see this directly from Eq.~(\ref{ssp1l}): 
if $k_c R_0 \equiv 1$ and if $p^{(1)}$ is convergent 
in the limit $\Lambda \rightarrow \infty$, then  
the power spectrum depends only on $k R_0$ 
and therefore obeys self-similarity. However, while Jain \& Bertschinger 
did show that leading infrared divergences cancel as loop momenta 
become vanishingly small, they did not compute loop corrections to the 
power spectrum explicitly. In fact, as we saw above, 
for $n=-1,-2$, $p^{(1)}$ develops a singularity at $k=k_c$, and the 
limit $\Lambda \rightarrow \infty$ does not exist. This necessarily 
introduces a dependence of the 1-loop power spectrum 
on $\Lambda=k_c/\epsilon$. As a result, even if 
$k_c R_0$ is held fixed, self-similarity is broken, 
{\it unless} 
$\epsilon$ is {\it also} assumed to scale with $R_0$ as $k_c$ does;  
we do not know of any arguments (analogous to that above) that would 
require $\epsilon$ to scale this way.  

We take a different viewpoint on this question. As Jain \& Bertschinger note,  
it is not surprising that choosing $k_c R_0 \approx 1$ throughout the 
evolution leads to self-similarity, since one is identifying 
the cutoff scale with the correlation length. In fact, this choice
can be characterized as artificially 
{\it introducing} self-similarity into the problem via the initial 
conditions. Here, we are 
rather interested in the question of whether self-similarity arises from  
the dynamical evolution over some range of 
scales far from the cutoffs 
(e.g., that given by Eq.~(\ref{ssreg})) when the initial 
conditions are not exactly scale-free. The basis of perturbation theory is 
the {\it assumption} 
that the statistical evolution of the large-scale fields can be determined 
perturbatively even when the density and velocity fields are highly 
non-linear on the smallest scales. Therefore, when $k_c R_0 > 1$, 
one should include 
perturbative contributions from scales below $R_0$ 
and see whether the results are sensible and/or self-similar, rather than 
continuously remove this power transfer by hand on the grounds that 
the theory is expected to  break down. 
This assumption of course {\it will} break 
down at late enough times and/or small enough scales, but it can be 
{\it tested}, and 
its range of validity quantified,  by checking its predictions 
against numerical simulations. {\it A fortiori}, 
our choice of fixed comoving cutoffs is
close to the conditions in (scale-free) N-body simulations, 
where infrared and ultraviolet cutoffs are
imposed by the box size and the grid scale, both of which are fixed in 
comoving coordinates. [This is not strictly true in the later stages 
of evolution for high-resolution codes (such as P$^3$M and tree codes), 
where the small-scale cutoff 
is eventually set by the mean interparticle distance; when the 
small-scale clustering stabilizes, this yields 
$k_c \propto a$, which, while not constant, has the {\it opposite} time 
dependence from the scaling needed for self-similarity, 
$k_c \propto a^{-2/(n+3)}$.] 

We expect that in models with relatively 
more small-scale power, i.e., larger $n$, 
the basic assumption of NLCPT will 
have a smaller range of validity: the power being fed from small 
to large scales soon invalidates the perturbative approach. In fact, the 
stronger breaking of self-similarity we have found in NLCPT at larger $n$ 
points to this conclusion, because scale-free N-body simulations 
exhibit self-similar scaling to very good accuracy for $n \ga -(1-3)$. 
To assess this issue quantitatively, 
we now turn to a comparison of NLCPT predictions with the ``universal'' 
scaling observed in scale-free numerical simulations.

\subsection{Comparison with Simulations: the Universal Scaling Hypothesis}
\label{sec:USH}

The universal scaling hypothesis~(\cite{HKLM91}) asserts that the non-linear 
evolution of the average two-point correlation function can be obtained 
from its linear counterpart by a universal scaling relation. This 
relation has been derived empirically 
from the study of numerical simulations for 
scale-free and CDM power spectra, but its main features can be 
understood on physical grounds~(\cite{NiPa94}). 
These results have been extended 
by Peacock \& Dodds (1994)
 to a universal relation between linear and non-linear power 
spectra and by Jain et al. (1995) 
to include the dependence of the scaling relation 
on the spectral index $n$. In its current version, the universal scaling 
hypothesis for the dimensionless power spectrum, 

\begin{equation}
	\Delta(k) \equiv 4 \pi k^3 P(k)
	\label{Delta},
\end{equation}

\noindent relates the linear ($\Delta_\ell$) to the non-linear 
evolved ($\Delta_e$) spectrum via 

\begin{equation}
	\Delta_e(k_e) = B(n) \ \Phi[ \Delta_\ell(k) / B(n) ],
	\ \ \ \ \ \ \ \ \ \ k^3 \equiv \frac{k_e^3}{1 + \Delta_e(k_e)}
	\label{USHps},
\end{equation}

\noindent where the simulations are empirically fit by setting 
$B(n) = (1 + n/3)^{1.3}$ and 

\begin{equation}
\Phi(z) = z \left( \frac{1+ 0.6 z +  z^2 - 0.2 z^3 - 1.5 z^{7/2} + z^4}
	{1 + 0.0037 z^3 } \right)^{1/2}
	\label{PhiUSH}.
\end{equation}

\noindent The 
ansatz in Eq.~(\ref{USHps}) is equivalent to the hypothesis that the mean 
dimensionless pairwise velocity is a universal function of the average 
two-point correlation function~(\cite{HKLM91,NiPa94}). Note that 
the linear spectrum is mapped to the non-linear spectrum at smaller 
scale ($k_e > k$), reflecting  
the fact that non-linearly evolving perturbations shrink in comoving 
coordinates. 
The specific relation between $k$ and $k_e$ in (\ref{USHps}) can be obtained 
from the evolution equation for the average correlation function 
(conservation of pairs), according to which  
the mean number of neighbors within distance $r_e$ of a 
particle is time independent if $r_e$ decreases in time according to
$r^3 \equiv r_e^3 [1 + \bar \xi_e(r_e)]$; here $r$ is a fixed scale,  
the radius of a sphere which encloses the same number of neighbors in the 
linear regime.

\begin{figure}[t!]
\centering
\centerline{\epsfxsize=10. truecm \epsfysize=9. truecm \epsfbox{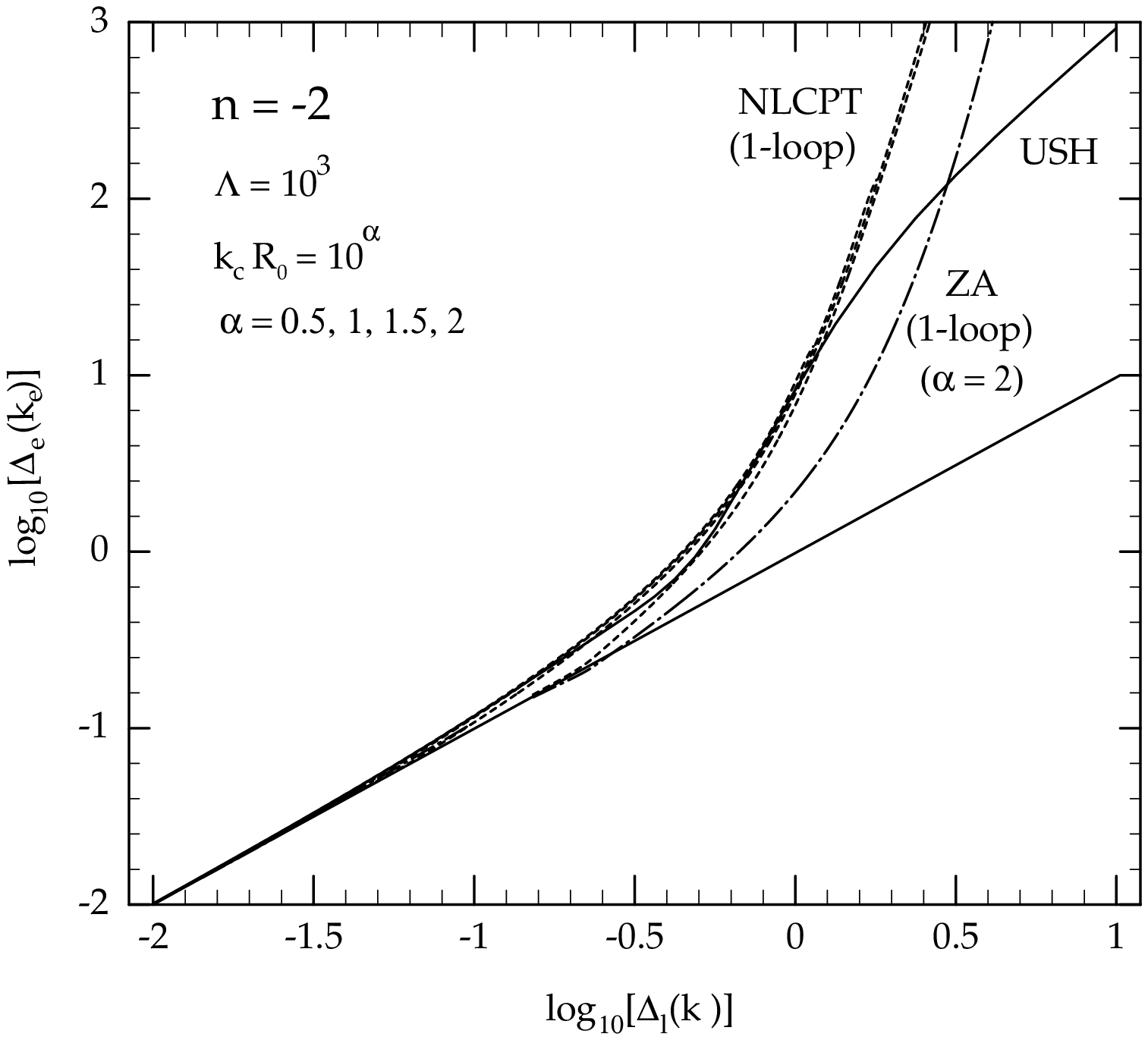}}

\caption{The evolved dimensionless power spectrum $\Delta_e(k_e)$ as a function
of the linear dimensionless power spectrum $\Delta_\ell(k)$
for $n=-2$. The scales $k_e$ and $k$ are related according to
Eq.~(\protect \ref{USHps}). The solid 
curve shows the universal scaling hypothesis (USH); dotted curves 
correspond to the predictions of non-linear cosmological perturbation
theory (NLCPT) including 1-loop corrections. The different NLCPT 
curves correspond to different values of $k_c R_0$ in the scaling 
region where self-similarity is expected to hold. The solid
straight line corresponds to linear perturbation theory, whereas
the dot-dashed curve
is the prediction of the 1-loop Zel'dovich approximation
for $k_c R_0 = 100$.}
\label{USHnm2}
\end{figure}

We now compare these expressions with the perturbative 
calculations of the previous sections. We can write the dimensionless 
power spectrum up to 1-loop corrections as (see Eq.~(\ref{ssp1l})):

\begin{equation}
	\Delta_e(k R_0,k_c R_0,\Lambda) \equiv \frac{(k R_0)^{n+3}}{ 
	\Delta I_\sigma(n,k_c R_0,\Lambda)}  + \frac{(k_c R_0)^{2n+3} 
	\ (k R_0)^{3}}{4 
	[\Delta I_\sigma(n,k_c R_0,\Lambda)]^2} \ p^{(1)}(n;k / k_c,
	\Lambda)
	\label{Delta1l}.
\end{equation}

\noindent In Fig.~\ref{USHnm2} we compare the 1-loop NLCPT relation 
between linear and non-linear dimensionless power 
spectra with the universal scaling hypothesis (USH) for spectral index $n=-2$. 
The different dotted 
curves correspond to the predictions of NLCPT for  
$k_c R_0=10^\alpha$, with $\alpha=0.5-2$; since $3 \la k_c R_0 \ll \Lambda 
=10^3$, cutoff effects in the linear evolution of the correlation 
length are negligible [see Eq.~(\ref{sscond}) and Fig.~(\ref{DInm2})]. 
Each curve extends  
down to scales such that $(k R_0)_{max} \equiv 
10^{\alpha-1/10}$.
The agreement between 
NLCPT and the USH (solid curve) is impressive all the way down to scales where 
$\Delta_e \approx 10$, far beyond the domain of linear perturbation theory
(see also \L okas et al. (1995) for a comparison between NLCPT and N-body 
results for $n=-2$).  
For smaller scales than this, the 1-loop power 
spectrum overestimates the non-linear power spectrum. Note, however, that 
the turnover at $\Delta_e \approx 10$ in the USH curve marks the onset of 
the transition to 
the regime described by the stable clustering hypothesis, and 
therefore disagreement with NLCPT beyond this point 
is not surprising. The NLCPT curve for 
$\alpha =2$ shows a slight disagreement with 
self-similarity at the large-scale end, which corresponds to the fact 
that, at this stage of evolution, these scales are not negligible 
compared to the size of the system. 
Otherwise, the agreement with self-similarity is excellent. 
For comparison, we also present the 
perturbative results to 1-loop for the Zel'dovich approximation (ZA) for $k_c 
R_0 = 100$ (dot-dashed curve).
The agreement between ZA and the USH is not very good, although 
the results of the N-body 
simulations of Jain et al. (1995)
 display a scatter comparable to the difference.  
In fact, their earlier (higher redshift) outputs seem to fit the  
ZA predictions shown here, suggesting 
that some of their scatter for $n=-2$ may be due to transients
from the ZA initial conditions of the simulations.

\begin{figure}[t!]
\centering
\centerline{\epsfxsize=10. truecm \epsfysize=9. truecm \epsfbox{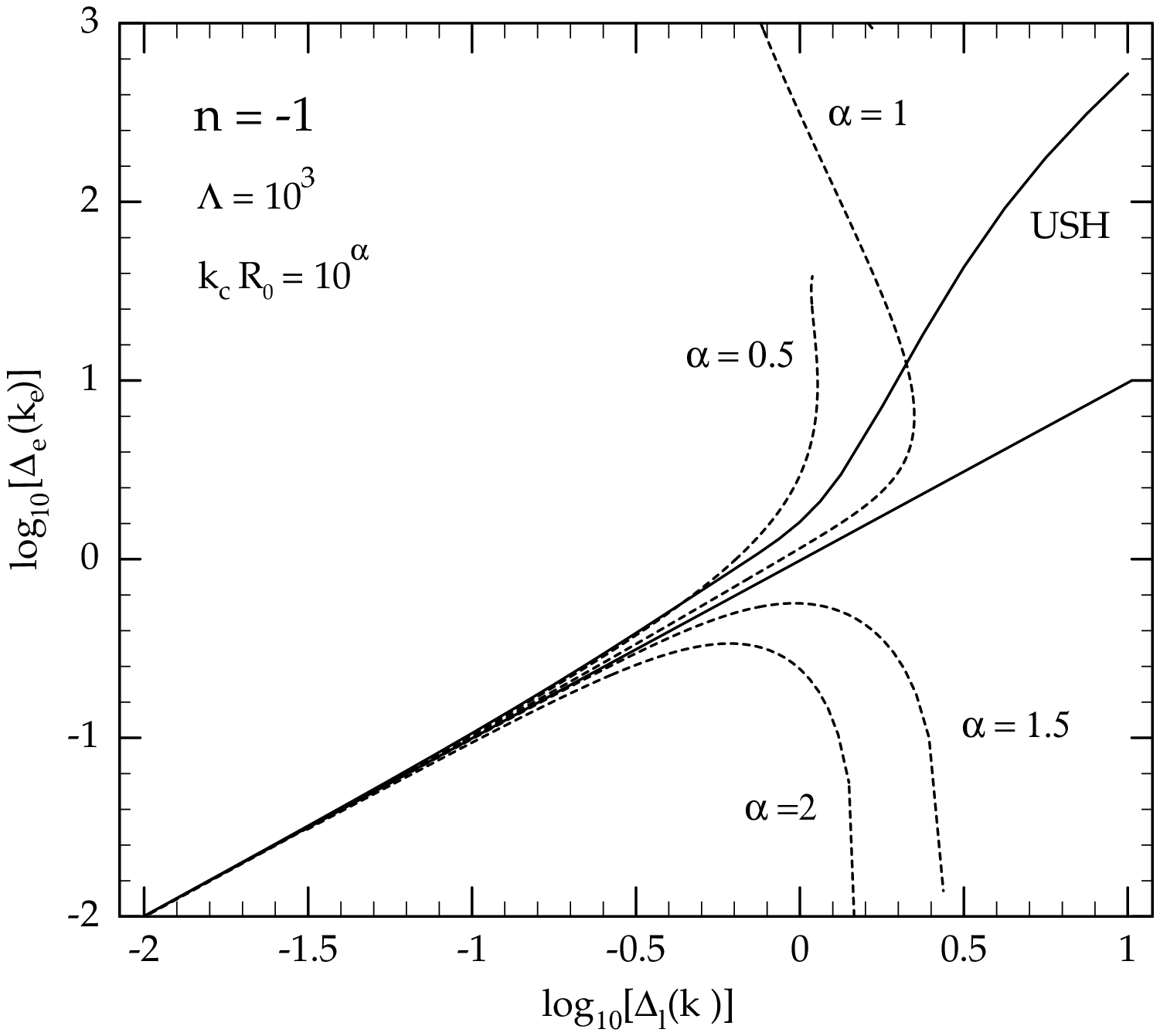}}

\caption{Same as Fig.~{\protect \ref{USHnm2}} but for $n=-1$. Note the
breaking of self-similarity.}
\label{USHnm1}
\end{figure}

\begin{figure}[t!]
\centering
\centerline{\epsfxsize=10. truecm \epsfysize=9. truecm \epsfbox{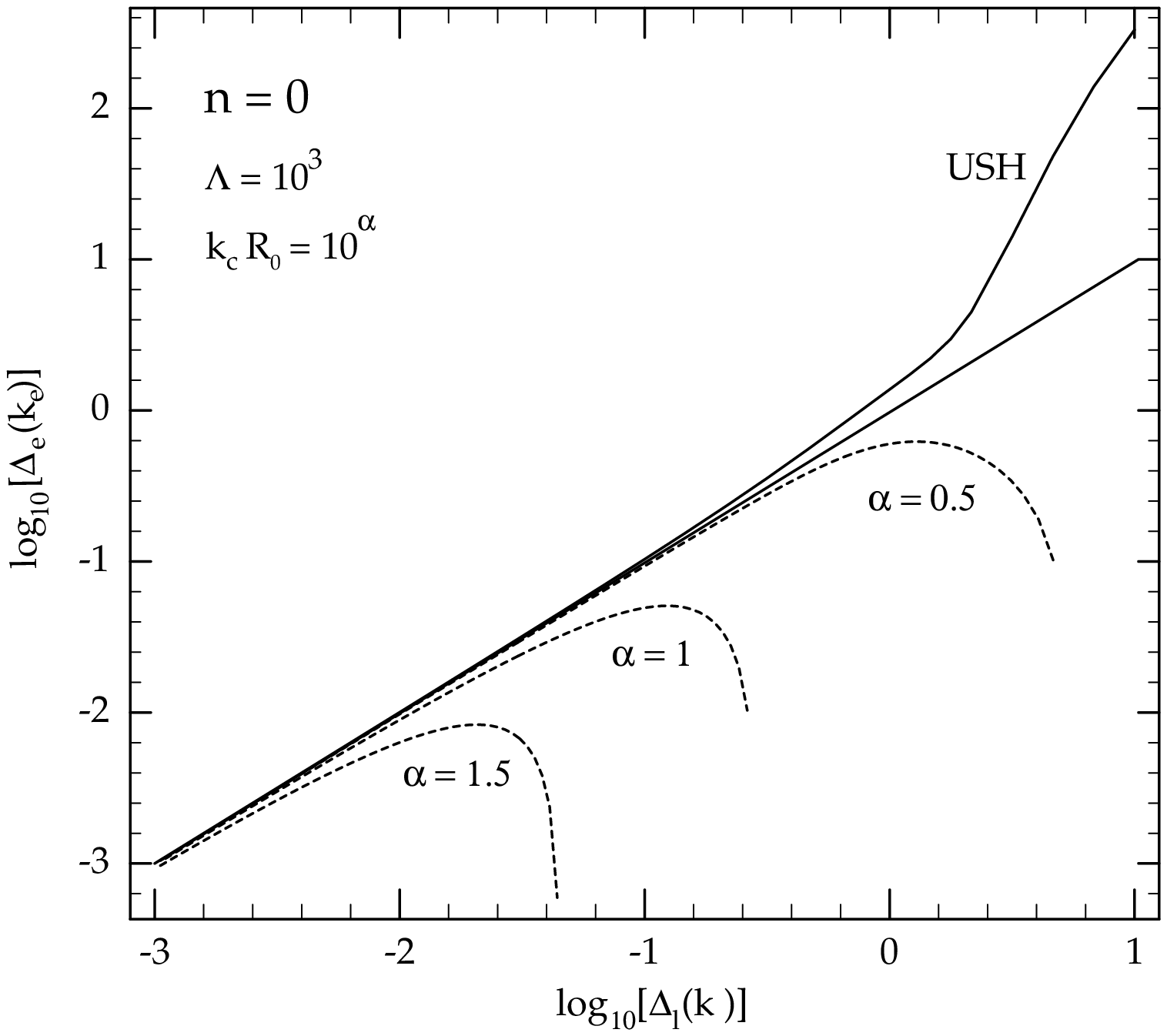}}

\caption{Same as Fig.~{\protect \ref{USHnm2}} but for $n=0$.}
\label{USHn0}
\end{figure}

\begin{figure}[t!]
\centering
\centerline{\epsfxsize=10. truecm \epsfysize=9. truecm \epsfbox{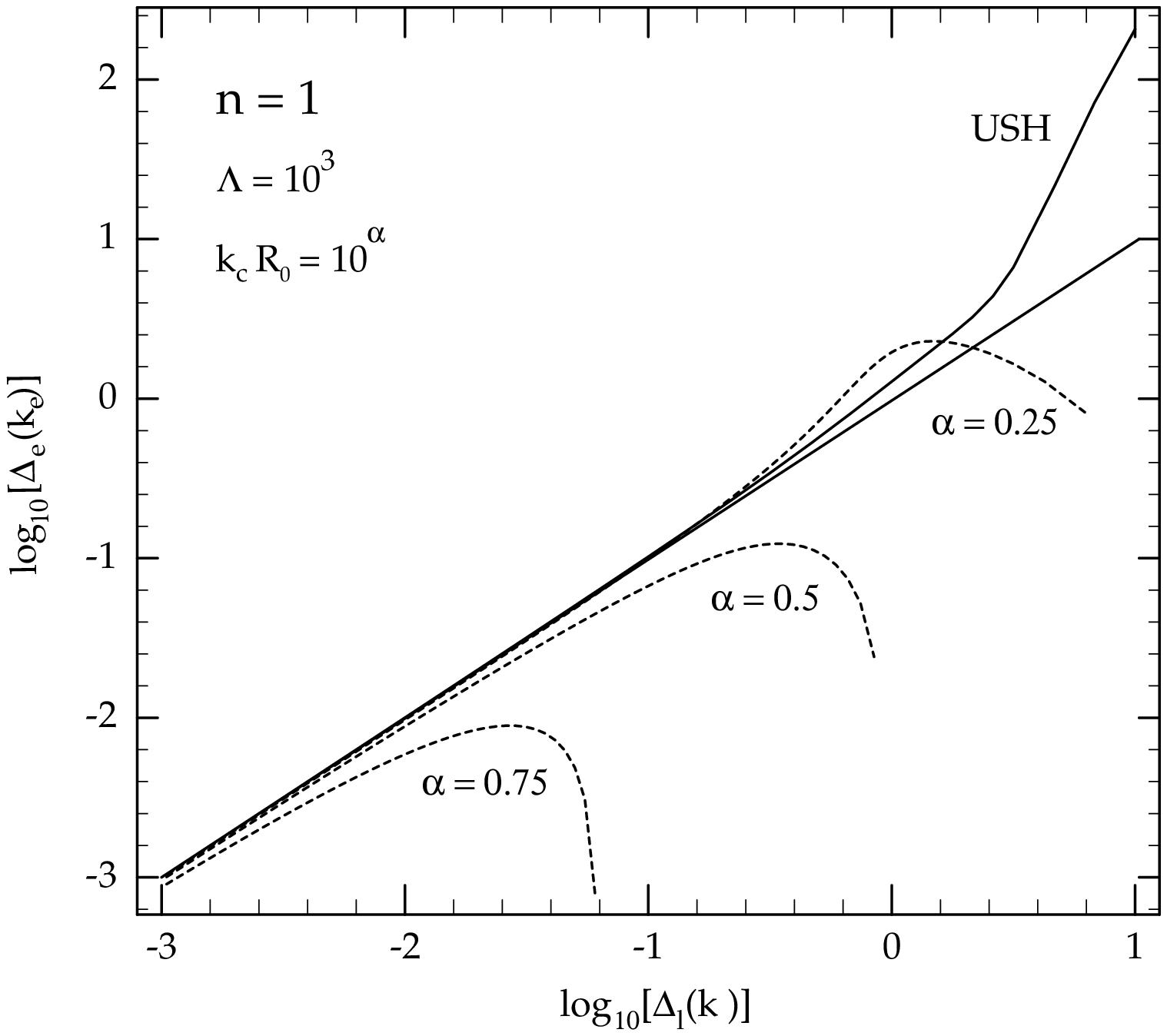}}

\caption{Same as Fig.~{\protect \ref{USHnm2}} but for $n=1$.}
\label{USHn1}
\end{figure}

In Figs.~\ref{USHnm1} - \ref{USHn1}, we show the same 
comparison for spectral indices $n=-1,0,1$. As expected 
from the discussion in the previous sections and the fact that the USH is 
self-similar by construction, for $n \geq -1$ there is obvious 
disagreement between NCLPT and USH. This discord 
becomes more evident as time 
evolution proceeds (a change in the scale factor by $\Delta a$ corresponds to  
$\Delta a \equiv 10^{(n+3)\Delta \alpha /3}$), as 
the scales considered decrease, or as the spectral 
index increases. The dotted curves that drop after reaching 
a maximum end 
close to the scale where the 1-loop corrections dominate over the 
tree-level contribution, driving  $\Delta_e$ negative (this happens 
at $k R_0 < 1$). 
A similar conclusion regarding the disagreement between NLCPT and the USH 
has been recently noted by Bharadwaj (1995), who examined the hypothesis 
that 
the mean dimensionless pairwise velocity is a universal function of the 
average correlation function. From the analysis of the BBGKY 
hierarchy in the weakly non-linear regime, he showed that this 
hypothesis is not obeyed at the 1-loop level
for linear power spectra $P_{11}(k) 
\propto k \exp (-k)$, consistent with our results for $n=1$.
In fact, we find that non-linear perturbation theory fares worse 
than linear theory for $n \ga -1$. 

A comparison of the  USH with NLCPT to 1-loop has also been carried out 
by \L okas et al. (1995), who concluded, based on numerical calculations of 
the $s^{(1)}$ 
coefficients for Gaussian smoothing, that the agreement was 
good. Their
main focus, however, was to understand whether non-linear evolution
decreases or increases the growth rate of structure with respect to that  
predicted by linear perturbation theory. In this regard, NLCPT 
qualitatively agrees with the USH in the sense that
both predict that the growth rates are increased over linear theory 
when $n=-2$ and decreased
when $n \geq 0$ (with $n=-1$ showing marginal behavior). However, as 
we have seen, the {\it quantitative} agreement is poor for $n \geq -1$: 
the scale-dependence of the 
$s^{(1)}$ coefficients in NLCPT disagrees with that of the  
USH, which by self-similarity requires  $s^{(1)} \approx {\rm const.}$ in
the scaling regime given by Eq.~(\ref{ssreg}). 

In fact, one can extract the 
$s^{(1)}$ coefficients predicted by the USH from Eqs.~(\ref{USHps}) 
and~(\ref{PhiUSH}) by using a small-$\Delta_e$ expansion,

 \begin{equation}
 	\Delta^{(1)} \approx \left[ 0.3 \ \left( \frac{3}{n+3}
 	\right)^{1.3} - \left( \frac{n+3}{3} \right) \right] \ \Delta_\ell^2
 	\label{delta1},
 \end{equation} 

\noindent [compare to Eq.~(\ref{Delta1l})]. 
Using Gaussian smoothing, this gives

 \begin{equation}
 	s^{(1)}_{\rm G}({\rm USH}) 
\approx \frac{2 \ \Gamma(n+3)}{\Gamma^2 \left( \frac{n+3}{2}\right)}
 	 \left[ 0.3 \ \left( \frac{3}{n+3}
 	\right)^{1.3} - \left( \frac{n+3}{3} \right) \right]
 	\label{s1USH},
 \end{equation} 

\noindent which yields $s^{(1)}_{\rm G} \approx 0.58, 
- 0.32, -3.56, -13.52$ for $n=-2,-1,0,1$ respectively. 
These results should be taken with caution, however, because  
the range over which the next-to-leading order term in the fitting formula 
(\ref{PhiUSH}) dominates is quite narrow~(\cite{JMW95}) and 
therefore sensitive to uncertainties in the fitting formulae. 
To quantify this uncertainty, one 
can, for example, calculate  $x^{(1)}_{\rm TH}$ 
for $n=-2$ from the result in Eq.~(\ref{delta1}), 
which gives $x^{(1)}_{\rm TH}(-2) \approx 0.50$; on the other hand, 
a more direct 
calculation from the fitting formula for the average correlation function 
(analogous to Eq.~(\ref{PhiUSH})) given in~\cite{JMW95} yields
$x^{(1)}_{\rm TH}(-2) \approx 0.75$, in remarkable agreement with the 
perturbative result, Eq.~(\ref{x1infnm2th}). 
For $n=-1,0,1$, the USH fitting formula for the 
average correlation function yields $x^{(1)}_{\rm TH} 
\approx -0.04, -0.55, -0.98$. We also 
note that these results and  Eq.~(\ref{s1USH}) 
predict a similar behavior regarding the change of sign of the 
1-loop corrections as $n$ increases, as pointed out by \L okas
et al. (1995).

\subsubsection{Self-similarity via Dimensional Regularization}
\label{sec:dimreg}

In the results presented so far, we imposed infrared and ultraviolet 
cutoffs on the initial power spectrum to obtain finite integrals. 
For $n=-2$, however, we have seen that 
the resulting 1-loop power spectrum 
in the scaling regime is independent of the cutoffs  
and thus self-similar; this suggests that in this case 
the loop corrections can be alternatively calculated without the need 
to employ cutoffs. In fact, 
if the spectral index is in the range
$-3 \leq n<-1$, one can show that 1-loop NLCPT preserves self-similarity 
by considering $p^{(1)}$ in the no-cutoff limit and using dimensional 
regularization to regularize the required integrations
(see Appendix B). Dimensional regularization is much simpler than 
the cutoff approach, because it obviates the need for complicated constraints 
on the angular integration variable [see Eq.~(\ref{lambdamm})]. It also 
makes possible the analytic calculation for non-integer values of $n$.

\begin{figure}[t!]
\centering
\centerline{\epsfxsize=10. truecm \epsfysize=9. truecm \epsfbox{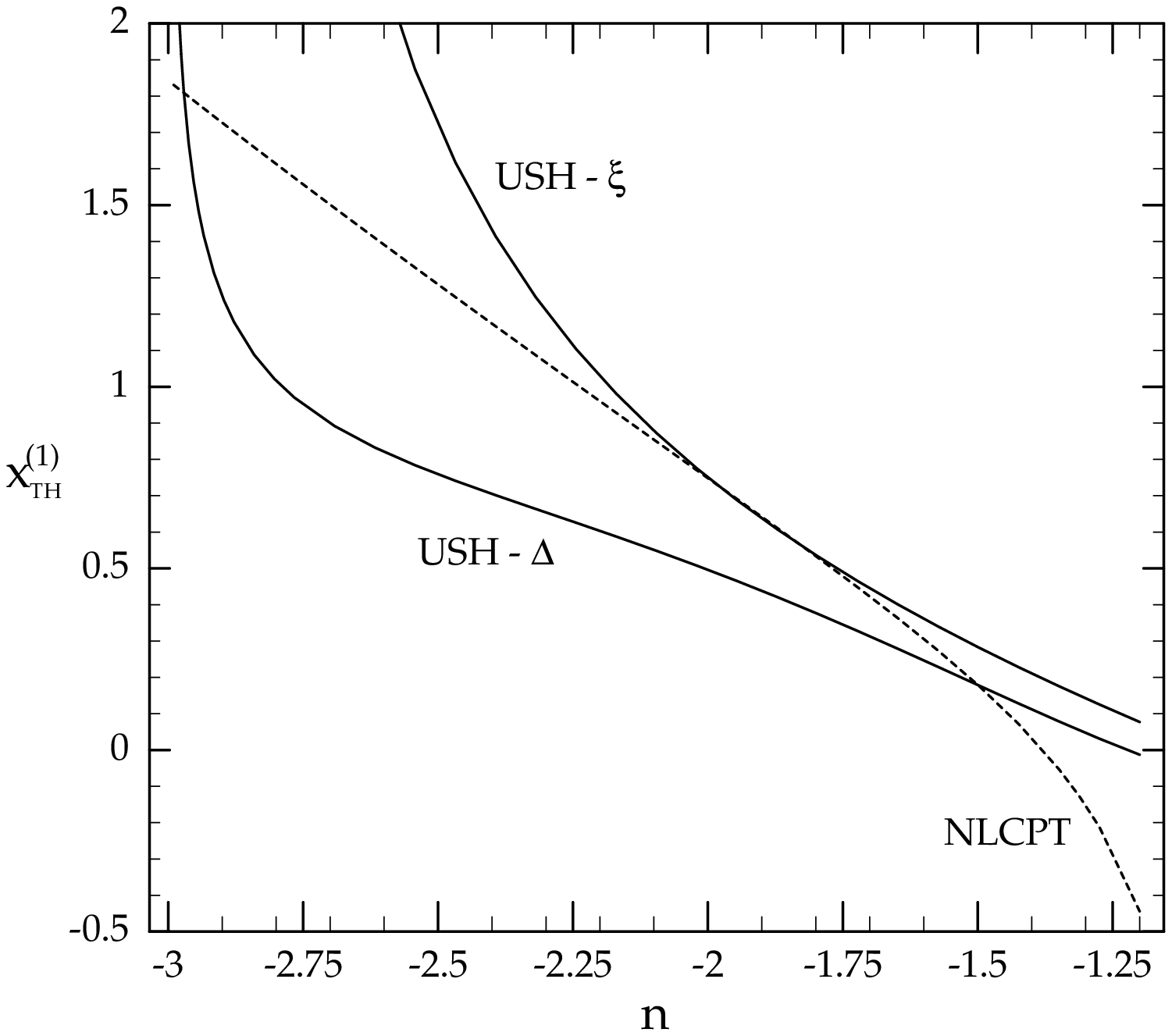}}

\caption{One-loop correction to the average correlation function, 
$x^{(1)}_{\rm TH}$,  
as a function of spectral index $n$ in NLCPT (dotted curve) and in the
USH obtained from the correlation function ansatz (solid curve labeled
USH - $\xi$) and from the dimensionless power spectrum ansatz 
(USH - $\Delta$).}
\label{x1DR}
\end{figure}

In Fig.~\ref{x1DR}, we 
show  $x^{(1)}_{\rm TH}$, derived via dimensional regularization    
for $-3 \leq n < -1$, and compare it with the 
predictions of the USH taken directly from the fitting formula for the 
average correlation function (labeled USH - $\xi$) and from the 
dimensionless power spectrum ansatz (USH - $\Delta$). We see that the 
agreement is very good over most of this range. 
The NLCPT results diverge as $n 
\rightarrow -1$, as expected from the logarithmic divergence found as $k_c
\rightarrow \infty $ in the cutoff calculation. As $n \rightarrow -3$, 
the NLCPT result 
goes to the unsmoothed value $x^{(1)} = s^{(1)} = 4063/2205 \approx 1.843$ 
~(\cite{ScFr96}); this can be understood from the fact that 
large scale fluctuations become dominant and averaging over small scales 
has a negligible effect.  (By contrast,  
the USH has a singularity at $n \rightarrow -3$ 
due to the way that the $n$-dependence is parametrized through $B(n)$ in
the USH fitting 
formula; since the USH was extracted from simulations with  
$n \geq -2$, it should probably only be trusted in that range.)
The linear part of the NLCPT 
curve is well described by an expansion about $n=-3$:

\begin{equation}
	x_{\rm TH}^{(1)} (n; \infty) = \frac{4063}{2205} - 
        \frac{3679}{2205} \ (n+3) \approx
	1.843 - 1.168 \ (n+3)
	\label{x1inftyTH}.
\end{equation}

\noindent Similar computations can be carried out in NLCPT for Gaussian 
smoothing and for the smoothed $s^{(1)}$ coefficient; 
the results  show essentially
the same features as displayed in Fig.~\ref{x1DR}.


\section{Summary and Conclusions}
\label{conclusions}

We have calculated the one-loop (first non-linear) 
corrections to the power spectrum, average
two-point correlation function, and variance of
the density field, including smoothing effects for top-hat and Gaussian
filters, for scale-free Gaussian initial conditions.  
For the power spectrum, these results should replace 
the expressions given in Makino et al. (1992) for the $p_{22}$ 
contribution on lengthscales smaller than the inverse 
ultraviolet cutoff of the linear 
spectrum ($k_c<k<2k_c$). Our results for the one-loop
corrections to the variance extend those of \L okas et al. (1995) to 
top-hat smoothing and correctly go over to the unsmoothed values found
in Scoccimarro \& Frieman (1996) when the smoothing radius approaches zero.

We found 
that, when formulated with ``fixed'' comoving cutoffs,  non-linear
perturbation theory beyond tree-level does not obey self-similar scaling 
when the spectral index $n \geq -1$. As a consequence, in this spectral 
range, NLCPT disagrees with the results of N-body simulations embodied in  
the self-similar universal scaling ansatz (USH). For $-3 < n < -1$, however, 
we found that one-loop 
corrections do obey self-similar scaling and are in excellent agreement
with the USH down to lengthscales where the dimensionless power spectrum 
$\Delta_e(k) \simeq \sigma^2(R\sim 1/k) 
\approx 10$; below this scale, the N-body results make the transition 
to the highly non-perturbative regime described approximately 
by stable clustering.

We interpret the breaking of self-similarity for $n \geq -1$
as a signature of the breakdown of the fundamental assumption that the 
large-scale evolution of the density field  can be calculated perturbatively
when there are highly non-linear fluctuations on small scales. In this 
instance, self-similarity breaking can be thought of as arising from 
an ultraviolet divergence
in the 1-loop corrections as the small-scale cutoff $k_c$ of the linear power
spectrum goes to infinity. This
problem is more severe as the spectral index $n$ increases, which is the
expected behavior due to the increase in small-scale power.

Since for $n \geq -1$, the relative 
power on small scales is large, the question arises of whether the 
disagreement between the perturbative results and  
numerical simulations is due to the 
breakdown of perturbation theory (as we argued above) or to the use of the 
single-stream approximation, which neglects the effects of pressure 
gradients due to velocity dispersion. 
When these effects are included, one expects  
that the anisotropic velocity dispersion (non-radial motions) 
associated with small-scale substructure can 
inhibit the collapse of large-scale perturbations---the ``previrialization'' 
effect (\cite{DP77,EvCr92,LJBH95,Peeb90}). In fact, as $n \rightarrow -1$ 
from below, 
$s^{(1)}$ changes sign and becomes negative, showing that  
such an effect is already present at some level 
in the single-stream approximation. In addition, 
the terms neglected in the single-stream  
approach have been included by Bharadwaj 
(1995), who nevertheless finds exactly the same results as in the 
single-stream approximation. Therefore, if velocity dispersion terms 
indeed become important, it appears that 
their effects cannot be treated perturbatively.
This suggests that the interpretation given 
above is the correct one, i.e., for $n \geq -1$ the non-linear 
evolution of the density power spectrum is inherently  
non-perturbative, and perturbative methods are of little use.

Given the excellent agreement between 1-loop NLCPT and N-body results 
for $n \la -1$, one is naturally led 
to ask whether this agreement can be improved upon, i.e., 
extended to smaller lengthscales, 
 by going to next order (2-loop
corrections). We regard this to be unlikely, since the USH
suggests that  the physical picture involved in this small-scale 
regime is well described
by stable clustering, which cannot be obtained perturbatively starting 
from linear
solutions. Another interesting question is whether the validity of loop 
corrections in NLCPT for higher order cumulants (such as 
the bispectrum, skewness, etc.) 
will follow the same pattern with spectral index. We will address this 
issue in future work.

\acknowledgments

We would like to thank S. Colombi, J. Fry, 
and R. Juszkiewicz for useful conversations. 
This research was supported in part 
by the DOE at Chicago and Fermilab 
and by NASA grant NAG5-2788 at Fermilab. After this paper was completed, we 
received a revised version of 
\cite{LJBH95}, which contains a number of improvements and corrections, 
made partly in response to our criticisms. 
They now agree with our numerical 
results for $s^{(1)}$ in the regime $k_c R \la 1$, 
and they have reproduced our analytic expressions for $s^{(1)}(n;k_c R 
\gg 1)$, our Eqns.(\ref{s1infnm2th} - \ref{s1infn1g}).

\appendix

\section{Expressions for $I_\sigma$ and $I_\xi$}
\label{app:IsigIxi}

Recall the definitions given in the text 
(see Eqs.~(\ref{Isigxi})) :

\begin{equation}
	I_\sigma (n, z) \equiv \int_0^z u^{n+2} \  W^2(u) \ du
	\label{Isigma},
\end{equation}

\begin{equation}
	I_\xi (n, z) \equiv \int_0^z u^{n+2} \  W(u) \ du
	\label{Ixi}.
\end{equation}

Here  
we tabulate these integrals as a function of filter and 
spectral index.

\subsection{Top-Hat Smoothing}
\label{app:TH}

\begin{mathletters}	 
\begin{eqnarray}
 	I_\sigma (-2, z) & = & \frac{3}{10 z^5} \Big[ -3 - 5 z^2 +
	(3 - z^2 + 2 z^4) \cos (2z) + (6 z + z^3) \sin (2 z) \nonumber \\ 
& & + 4 z^5 {\rm Si}(2z) \Big]
	\label{Isinm2TH},  \\
	I_\sigma (-1, z) & = & \frac{9}{4} + \frac{9}{8 z^4} \Big[-1 - 
	2 z^2 + \cos (2z) + 2 z \sin (2z) \Big]
	\label{Isinm1TH},  \\
	I_\sigma (0, z) & = &\frac{3}{2 z^3} [ -1 -3 z^2 + 
	(1 + z^2 ) \cos (2z) + 2 z \sin (2z) + 2 z^3 {\rm Si}(2z) ]
	\label{Isin0TH},  \\
	I_\sigma (1, z) & = & -\frac{9}{2} ( 1 - \gamma_e - \ln 2 ) +
	\frac{9}{4 z^2} [-1 + \cos (2z) - 2 z^2 {\rm Si}(2z) + 2 z^2 \ln (z) 
\nonumber \\ & & + 2 z \sin (2z) ]
	\label{Isin1TH},
\end{eqnarray}
\end{mathletters}	 

\noindent where ${\rm Si}(z)$ and ${\rm Ci}(z)$ denote the sine 
and cosine integrals, defined by

\begin{eqnarray}
	{\rm Si}(z) & \equiv  & \int_0^z \frac{\sin (u)}{u} du
	\label{Si},  \\
	{\rm Ci}(z) & \equiv & - \int_z^\infty \frac{\cos (u)}{u} du
	\label{Ci},
\end{eqnarray}

\noindent and $\gamma_e \simeq  0.577216...$ is  
the Euler-Mascheroni constant. As $z \rightarrow \infty$, we have ${\rm 
Ci}(z) \rightarrow 0$ and ${\rm Si}(z) \rightarrow \pi /2$. Using the 
following  expression of a top-hat filter in terms of Bessel functions:

\begin{equation}
	W_{\rm TH} (u) = 3 \sqrt{\frac{\pi}{2}} u^{-3/2}  J_{3/2}(u)
	\label{tophatbessel},
\end{equation}¥

\noindent we have (for any $n$ in the range $-3 < n < 1$):

\begin{equation}
	I_\sigma (n, \infty) = 9 \pi 2^{n-2} \frac{ \Gamma (1-n) \Gamma 
	[(n+3)/2] }{\Gamma^2 (1-n/2) \Gamma [(5-n)/2]}
	\label{IsiginftyTH},
\end{equation}

\noindent where $\Gamma(x)$ denotes the gamma function.
For $I_\xi (n, z)$ we obtain:

 \begin{mathletters}
 
 \begin{equation}
 	 I_\xi (-2, z)  =  \frac{3}{2} \left[ {\rm Si}(z) + 
	\frac{\cos (z)}{z} - \frac{\sin (z)}{z^2} \right]
	\label{Ixinm2TH},
 \end{equation}
 	 
\begin{equation}
 	 	 	I_\xi (-1, z)  =  3 [ 1 - \sin (z) /z ]
	\label{Ixinm1TH},
\end{equation}

\begin{equation}
 	 	 	I_\xi (0, z)  =  3 [ {\rm Si}(z)  - \sin (z) ]
	\label{Ixin0TH},
\end{equation}

\begin{equation}
 	 	 	I_\xi (1, z)  =  3 [ 2- 2 \cos (z)  - z \sin (z) ]
	\label{Ixin1TH}.
\end{equation}

 \end{mathletters}	
 
 \noindent Using Eq.~(\ref{tophatbessel}), we have ($-3<n<0$):
 
 \begin{equation}
	I_\xi (n, \infty) = 3 \sqrt{\pi} 2^{n} \frac{  \Gamma 
	[(n+3)/2] }{\Gamma (1-n/2) }
	\label{IxiinftyTH}.
\end{equation}

\subsection{Gaussian Smoothing}
\label{app:G}

 \begin{mathletters}	 
\begin{eqnarray}
	I_\sigma (-2, z) & = & \frac{\sqrt{\pi}}{2} {\rm erf}(z) 
	\label{Isinm2G}  \\
	I_\sigma (-1, z) & = & \frac{1}{2} [1 - \exp(-z^2)]
	\label{Isinm1G}  \\
	I_\sigma (0, z) & = & \frac{\sqrt{\pi}}{4} \ {\rm erf}(z) -
	\frac{1}{2} z \exp(-z^2)
	\label{Isin0G}  \\
	I_\sigma (1, z) & = & \frac{1}{2} - \frac{(1 + z^2 )}{2} \exp(-z^2)
	\label{Isin1G}
\end{eqnarray}
 \end{mathletters}	 
 
\noindent where ${\rm erf}(z) $ denotes the error function, 
\begin{equation}
	{\rm erf}(z)  \equiv \frac{2}{\sqrt{\pi}} \int_0^z \exp (-u^2) du
	\label{erf},
\end{equation}

\noindent and for $I_\xi (n, z)$ we have:

\begin{equation}
	I_\xi (n, z) = 2^{(n+3)/2} \ I_\sigma (n, z/\sqrt{2})
	\label{Ixi-Isi}.
\end{equation}

\noindent When $z \rightarrow \infty$ we obtain:

\begin{equation}
	I_\sigma (n, \infty) = \frac{1}{2} \ \Gamma \left( \frac{n+3}{2} 
\right)
	\label{IsiginftyG}.
\end{equation}

\section{Dimensional Regularization}

To obtain the low-$k$ behavior of the power spectrum for $n <-1$, one
can 
use dimensional regularization to simplify considerably the calculations, and
get one-loop coefficients such as $x^{(1)}(n;\infty)$ for $n$ in the  
range $-3 < n < -1$. Since we are interested in the limit 
$k_c \rightarrow \infty$,
all the integrals run from $0$ to $\infty$, and  divergences are 
regulated by changing the dimensionality $d$ of space: 
we take $d = 3 + \epsilon$ and expand 
in $\epsilon \ll 1$. For power spectrum calculations, we need  
the following standard formula for dimensional-regularized 
integrals~(\cite{Collins84}):

\begin{equation}
\int  \frac{d^d {\bf q}}{(q^2)^{\nu_1} [({\bf k}-{\bf q})^2]^{\nu_2}} =
 \frac{\Gamma (d/2 -\nu_1)
\Gamma (d/2 -\nu_2)
\Gamma (\nu_1 + \nu_2 - d/2)}{\Gamma (\nu_1) \Gamma (\nu_2) \Gamma (d-
\nu_1 - \nu_2)} \ \pi^{d/2}  \ k^{d-2 \nu_1 -2 \nu_2}
\label{den2}.
\end{equation}

\noindent When using this equation, divergences appear as poles in the 
gamma functions, which can be handled by  the following expansion 
($n=0,1,2, \ldots$ and $\epsilon \rightarrow 0$):

\begin{equation}
	\Gamma(-n + \epsilon) = \frac{(-1)^n}{n!} \left[ \frac{1}{\epsilon} + 
	\psi (n+1) + \frac{\epsilon}{2} \left( \frac{\pi^2}{3} + \psi^2 (n+1) - 
	\psi' (n+1) \right) + {\cal O} (\epsilon^2) \right]
	\label{gammapoles},
\end{equation}

\noindent where $\psi (x) \equiv d \ln \Gamma (x) /dx$ and

\begin{mathletters}
\begin{eqnarray}
	\psi (n+1) & = & 1 + \frac{1}{2} + \ldots +  \frac{1}{n} - \gamma_e
	\label{psi},  \\
	\psi' (n+1) & = & \frac{\pi^2}{6} - \sum_{k=1}^{n} \frac{1}{k^2}
	\label{psiprime},
\end{eqnarray}
\end{mathletters}

\noindent with $\psi (1) = - \gamma_e$ and $\psi' (1) = \pi^2 /6$.
We can write the one-loop power spectrum as
(see Eq.~(\ref{P1lamp}) ):

\begin{equation}
         P^{(1)}(k,\tau ; n)  =   A^2 a^4 (\tau)  
\int d^d {\bf q} \ q^n \left( 2 |{\bf k}-{\bf q}|^n  \
[F_2^{(s)}({\bf k}-{\bf q},{\bf q}) ]^2    + 6  k^n  \
         F_3^{(s)}({\bf k},{\bf q},-{\bf q}) \right)
        \label{P1DR}. 
\end{equation}

\noindent In the numerators of the integrands, we can use the relation  
$2 \ {\bf q} \cdot {\bf k} = - ({\bf k}-{\bf q})^2
+ k^2 + q^2$ to rewrite Eq.~(\ref{P1DR}) exclusively in terms of integrals
of the form~(\ref{den2}).
The resulting 1-loop power spectrum contributions for $-3 < n < 
-1$ are 

\begin{eqnarray}
        P_{22}(k,\tau ; n) & =&  \Bigg(
\frac{\Gamma (5/2 -n) \Gamma^2 [(n-1)/2]}{2 \Gamma^2 (2-n/2) \Gamma (n-1)} +
\frac{3 \Gamma (3/2 -n) \Gamma [(n-1)/2] \Gamma [(n+1)/2]}{
 \Gamma (1-n/2) \Gamma (2-n/2) \Gamma (n)} \ \ \ \ \ \ \ \ \ \ 
 \nonumber \\ & & +  
\frac{29 \Gamma (1/2 -n) \Gamma^2 [(n+1)/2]}{4 \Gamma^2[1-n/2] \Gamma (n+1)}
+  
\frac{11 \Gamma (1/2 -n) \Gamma [(n-1)/2] \Gamma [(n+3)/2]}{
4 \Gamma (2-n/2) \Gamma ( -n/2) \Gamma (n+1)} \nonumber \\ & & - 
 \frac{15 \Gamma (-1/2 -n) \Gamma [(n-1)/2] \Gamma [(n+5)/2]}{
2 \Gamma (-1-n/2) \Gamma (2 -n/2) \Gamma (n+2)} + 
\frac{15 \Gamma (-1/2 -n) \Gamma [(n+1)/2]}{
2 \Gamma (1-n/2) \Gamma (-n/2)} \nonumber \\ & &
\times \frac{ \Gamma [(n+3)/2]}{ \Gamma (n+2)} -  
 \frac{25 \Gamma (-3/2 -n) \Gamma [(n+1)/2] \Gamma [(n+5)/2]}{
 \Gamma (-1-n/2) \Gamma (1-n/2) \Gamma (n+3)} + 
\frac{25 \Gamma (-3/2 -n) }{
4 \Gamma (-2-n/2)} \nonumber \\ & &
\times \frac{ \Gamma [(n-1)/2] \Gamma [(n+7)/2]}{ \Gamma (2-n/2) \Gamma (n+3)}
+  \frac{75 \Gamma (-3/2 -n) \Gamma^2 [(n+3)/2]}{4 \Gamma^2[-n/2] \Gamma (n+3)}
 \Bigg) \ \frac{\pi^{3/2}  A^2 a^4 (\tau)}{49} \nonumber \\
& & \times  k^{2n+3}, 
\label{P22DR}
\end{eqnarray}

\begin{eqnarray}
         P_{13}(k,\tau ; n) & = & \Bigg(
- \frac{\Gamma [(n+1)/2] \Gamma [(1-n)/2]}{84 \Gamma (1-n/2) \Gamma (1+n/2)}
- \frac{19 \Gamma [-(n+3)/2] \Gamma [(n+5)/2] }{84 
 \Gamma (-1-n/2) \Gamma (3+n/2) } \nonumber \\ & & +
\frac{ \Gamma [-(n+5)/2] \Gamma [(n+7)/2]}{12 \Gamma[-2-n/2] \Gamma (4+n/2)}
+ 
\frac{5  \Gamma [-(n+1)/2] \Gamma [(n+3)/2]}{
28 \Gamma (2+n/2) \Gamma ( -n/2) } \nonumber \\ &  & -
 \frac{  \Gamma [(n-1)/2] \Gamma [(3-n)/2]}{
42 \Gamma (2-n/2) \Gamma (n/2) }  \Bigg) \ \pi^{2}  A^2 a^4 (\tau) \
k^{2n+3}. 
\label{P13DR}
\end{eqnarray}

\noindent Note that this implies that  
$P^{(1)} \propto k^{2n+3}$, as required by self-similarity.



\begin{thebibliography}{ZZZZZZZZZZZ1999}

\bibitem[Barenblatt 1979]{Barenblatt79}Barenblatt, G. I. 1979, 
{\it Similarity, Self-Similarity, and Intermediate Asymptotics}, 
Consultants Bureau, New York (1979).
	
\bibitem[Baugh \& Efstathiou 1994]{BaEf94}Baugh, C. M. \&  Efstathiou, G. 
1994, MNRAS, 270, 183
	
\bibitem[Baugh et al. 1995]{BGE95}Baugh, C. M., Gazta\~{n}aga, E., 
\&  Efstathiou, G. 1995, MNRAS 274, 1049 

\bibitem[Bernardeau 1994]{Bernardeau94c}Bernardeau, F. 1994, 
A\&A 291, 697 

\bibitem[Bertschinger \& Gelb 1991]{BG91}Bertschinger, E. \& Gelb, J. 1991,
Comp. in Physics, 5, 164

\bibitem[Bharadwaj 1995]{Bhar95}Bharadwaj, S. 1995, preprint, 
astro-ph/9511085 

\bibitem[Coles 1990]{Coles90}Coles, P. 1990, MNRAS, 243, 171

\bibitem[Colombi et al. 1995]{CBH95}Colombi, S., Bouchet, F. R., \& 
Hernquist, L. 1995, preprint, astro-ph/9508142 

\bibitem[Collins 1984]{Collins84}Collins, J. C. 1984, {\it Renormalization}, 
(Cambridge: Cambridge University Press) 


\bibitem[Davis \& Peebles 1977]{DP77}Davis, M. \& Peebles, P. J. E. 1977,
ApJS, 34, 25

\bibitem[Efstathiou et al. 1988]{EFAL88}Efstathiou, G., Frenk, C. S., 
White, S. D. M., \& Davis, M. 1988, MNRAS, 235, 715

\bibitem[Evrard \& Crone 1992]{EvCr92}Evrard A. E. \& Crone M. M. 1992, 
ApJ, 394, L1

\bibitem[Fry 1984]{Fry84}Fry, J. N. 1984, ApJ, 279, 499

\bibitem[Fry 1994]{Fry94}Fry, J. N. 1994, ApJ, 412, 21
	
\bibitem[Gazta\~{n}aga \& Baugh 1995]{GaBa95}Gazta\~{n}aga, E. \& Baugh, 
C. M. 1995, MNRAS 273, L1

\bibitem[Goldenfeld 1992]{Goldenfeld92}Goldenfeld, N. 1992, 
{\it Lectures on Phase 
Transitions and the Renormalization Group}, Addison Wesley (1992).

\bibitem[Goroff et al. 1986]{GGRW86}Goroff, M. H., Grinstein, B., 
Rey, S.-J., \& Wise, M. B. 1986, ApJ, 311, 6 

\bibitem[Gramann 1992]{G92}Gramann, M. 1992, ApJ, 401, 19

\bibitem[Hamilton et al. 1991]{HKLM91}Hamilton, A. J. S., Kumar, P., Lu, E.,  
\& Matthews, A. 1991, ApJ , 374, L1


\bibitem[Jain \& Bertschinger 1994]{JaBe94}Jain, B. \& Bertschinger, E. 1994,  ApJ, 431, 495 

\bibitem[Jain \& Bertschinger 1995]{JaBe95}Jain, B. \& Bertschinger, E. 1995, 
preprint, astro-ph/9503025

\bibitem[Jain et al. 1995]{JMW95}Jain, B.,  Mo, H. J., \& White, S. D. M. 1995, 
MNRAS, 276, L25 

\bibitem[Jain  1995]{Jain95}Jain, B.  1995, 
preprint, astro-ph/9509033

\bibitem[Juszkiewicz 1981]{Juszkiewicz81}Juszkiewicz,  R. 1981, MNRAS, 197, 931

\bibitem[Juszkiewicz et al. 1993]{JBC93}Juszkiewicz, R., Bouchet, F. R., \&
Colombi, S. 1993, ApJ, 412, L9

\bibitem[Juszkiewicz et al. 1984]{JSB84}Juszkiewicz, R., Sonoda, D. H., \&  Barrow, J. D. 1984, MNRAS, 209, 139  

\bibitem[Lewin 1981]{Lewin81}Lewin, L. 1981, {\it Polylogarithms and Associated 
	Functions}, North Holland 

\bibitem[\L okas et al. 1995]{LJBH95}\L okas, E. L., Juszkiewicz, R., Bouchet, F. 
R., \& Hivon, E. 1995, preprint, astro-ph/9508032 


\bibitem[Makino et al. 1992]{MSS92}Makino, N., Sasaki, M., \& and Suto, Y. 1992, 
Phys. Rev. D, 46, 585 

\bibitem[Nityananda \& Padmanabhan 1994]{NiPa94}Nityananda, R. \&  
Padmanabhan, T. 1994, MNRAS, 271, 976

\bibitem[Padmanabhan et al. 1995]{PCOS95}Padmanabhan, T., Cen R., 
Ostriker J. P. \& Summers F. J. 1995, preprint, astro-ph/9506051

\bibitem[Peacock \& Dodds 1994]{PeDo94}Peacock, J. A. \& Dodds, S. J. 1994, 
MNRAS, 267, 1020 

\bibitem[Peebles 1974]{Peebles74}Peebles, P. J. E. 1974, A\&A, 32, 391 	

\bibitem[Peebles 1980]{Peebles80}Peebles, P. J. E. 1980, 
{\it The Large-Scale Structure of the Universe}, Princeton University Press 

\bibitem[Peebles 1990]{Peeb90}Peebles, P. J. E. 1990, ApJ, 365, 27
	
\bibitem[Peebles \& Groth 1976]{PeGr76}Peebles, P. J. E. \& Groth, 
E. J. 1976, A\&A, 53, 131 

\bibitem[Ryden \& Gramann 1991]{RG91}Ryden, B. \& Gramann, M. 1991, ApJ, 
383, L33
	
\bibitem[Scoccimarro \& Frieman 1996]{ScFr96}Scoccimarro, R. \& Frieman, J. 
1996,  ApJS, in press 

\bibitem[Suto \& Sasaki 1991]{SuSa91}Suto, Y., \& Sasaki, M. 1991, 
Phys. Rev. Lett., 66, 264 

\bibitem[Vishniac 1983]{Vishniac83}Vishniac, E. T. 1983, MNRAS, 203, 345

\bibitem[Wise 1988]{Wise88}Wise, M. B. 1988, in {\it The Early Universe},
 W.G. Unruh and G. W. Semenoff (eds.), D. Reidel Publishing Company

\bibitem[Zel'dovich 1965]{Zel'dovich65}
Zel'dovich, Ya. B. 1965, Adv. Astron. Ap., 3, 241 

\end{thebibliography}
\end{document}